\documentclass[12pt,english]{article}
\usepackage[T1]{fontenc}
\usepackage[latin9]{inputenc}
\usepackage{babel}
\usepackage{array}
\usepackage{float}
\usepackage{multirow}
\usepackage{amsmath}
\usepackage{amssymb}
\usepackage{graphicx}
\usepackage{esint}
\usepackage[unicode=true,
 bookmarks=false,
 breaklinks=false,pdfborder={0 0 1},backref=section,colorlinks=false]
 {hyperref}

\makeatletter

\providecommand{\tabularnewline}{\\}

\newcommand{\fig}[1]{Fig. \ref{#1}}
\newcommand{\tabl}[1]{Table \ref{#1}}

\makeatother

\makeatother

\begin{document}
\title{Fractal-cluster theory and thermodynamic principles of the control
and analysis for the self-organizing systems }
\author{V.T. Volov \thanks{Department of Natural Sciences, Samara State University of Railway
Transport, Samara, Russia, e-mail: volovvt@mail.ru}}
\date{\today}

\maketitle

\begin{abstract}
The theory of resource distribution in self-organizing systems on
the basis of the fractal-cluster method has been presented. In turn,
the fractal-cluster method is based on the fractal-cluster relations
of V.P. Burdakov and the analytical apparatus of the thermodynamics
of I. Prigozhin's structure. This theory consists of two parts: deterministic
and probabilistic. The first part includes the static and dynamic
criteria, the fractal-cluster dynamic equations which are based on
the Fibonacci's range characteristics fractal-cluster correlations.
The second part includes fundamentals of the probabilistic theory
of a fractal-cluster systems. This part includes the dynamic equations
of the probabilistic evolution of these systems. By using the numerical
researches of these equations for the stationary case the random state
field of the self-organizing system in the phase space of the $D$,
$H$ criteria and probability $f$ have been obtained. For the socio-economical
and biological systems this theory has been tested. In particular,
three fundamental fractal-cluster laws have been obtained for biological
organisms: probabilistic, energetic and evolutionary.
\end{abstract}
\tableofcontents{}
\bigskip{}
\bigskip{}

\section{Introduction}

{}\label{Introduction}

{}The application of natural science and mathematical methods in
biological and socio-economical systems science has a long history.
Beginning with the XVI-XVII centuries there was an active introduction
of analytical tools of physics and thermodynamics in almost all sections
of biological science, which subsequently led to the birth of biophysics
and biological thermodynamics. The fundamental works that laid the
foundations for the main directions in the development of the synthesis
of two sciences -- biology and physics -- are the classical studies
of the famous natural scientists W. Harvey, G. A. Borelli, L. Euler,
M. Lomonosov, A. L. de Lavoisier, L. Galvani, A. Volta, J. Banks,
J. Bernstein, H. Helmholtz and others. The XX century was marked by
a further process of convergence of biology with phenomenological
thermodynamics, which is associated with the names of such famous
scientists as W. Nernst, T. Hill, A. Hodgkin, A. Huxley, J. Eccles
and many others. The end of the XX century and the beginning of the
XXI century were characterized by the next convergence of biology
with the nonequilibrium thermodynamics by I. Prigogine \cite{PN,GP}.
In the 21st century, this process continued in the studies of various
groups and schools, among which we would like to highlight the works
of A. I. Zotin, A. Alimov, T. Kazantseva, A. A. Zotin, von Stockar
and others \cite{Westerhoff,VonStockar,VonStockar1,Liu,VonStockar2,VonStockar3,Alimov1,Alimov,Zotin2,Zotin3,Zotin1,Zotin},
as they overlap to some extent with our research.

{}During the last few decades the mathematical modeling methods of
physical systems have been actively used for the description of the
different systems of non-physical nature, for example, socio-economical
systems. On the boundary of such scientific areas as statistic physics,
random processes theory, nonlinear dynamics on the one hand and macro-and
microeconomics on the other hand there has been created a new interdisciplinary
scientific area which has received the name ``econophysics'' \cite{PN}.
The term ``econophysics'' had been suggested by an American physicist,
H. Stanley, in 1995 for integration of numeral researches which used
typical physical methods for the solution of social and economic tests
\cite{GP}.

{}Obviously the first attempt of the physical method modeling using
the economic processes we can assume a research of the French mathemation
L. Bacheliar \cite{Bach}. L. Bacheliar made an attempt to descute
the financial range dynamics by using an analogy with random physical
processus -- Brownian movement. Later in 1965 B. Mandelbrot revealed
that financial range dynamics is absolutely the same for small and
large time dimensions \cite{Mand}. This quality B. Mandelbrot named
``self-similarity'' and objects have the same quality -- ``a fractal''.
Physics models which have been applied in sociology and economics
include the kinetic theory of gas, ferromagnetics, percolation models,
chaotic dynamics models and other. Moreover, there have been attempts
to use the mathematical theory of complexity and information theory.
Since socio-economic phenomena are the result of the interaction among
many heterogeneous objects, there is an analogy with statistical mechanics
and thermodynamics. There are also attempts to use the physical ideas
from such areas as fluid dynamics, classical and quantum mechanics
and the path integral formulation of statistical mechanics.

{}A special place among numerous researches devoted to the study
of biological systems and organisms on the basis of natural and mathematical
tools is occupied by the cycle of works connected with the application
of fractal geometry in this field \cite{St}. For example, Burdakov's
study may be related to these researches \cite{Burd}. A brief historical
excursion in the creation and development of the fractal-cluster approach
for the study of biological organisms and socio-economical systems
consists in the following. In the 60's of the previous century V.
P. Burdakov was a young employee of a famous firm ``Energy'' --
flagship of Russian cosmonautics. The general director, the academician
S. Korolev, proposed to his young employee the problem to determine
the optimal ratio of rocket resources related to its energy, transport,
security, technology and control units. For the statistic analysis
of complex self-organizing systems (SOS) V. B. Burdakov used extensive
parameters (mass, volume, time, financial resource) instead of intensive
thermodynamic parameters (temperature, pressure). First of all he
analyzed the best Russian and American rockets which were independently
created by a number of scientists and engineers from different countries.
Of course, rockets are the most complex technical systems which can
be considered as self-organizing systems in the ``man--technical
system'' class. The shares of resource (mass) of a rocket which refer
to energy, transport, security, technology and control subsystems
were determined. Later, when analysing organismal systems, the security
and control subsystems were renamed into the ecology and information
subsystems. These shares in the best Russian and American rockets
turned out to be approximately equal. Besides, the corresponding resource
distributions of various technical systems (the ``man--technical
system'' class) were also studied. The next step of his investigation
was the study of biological organisms which were created by nature
as a result of evolution. He determined the relations of the mass
shares of organisms (39 organism species starting with Chlamydomonas
and ending with whales) with their energy, transport, ecology, technology
and information needs. These basic needs have been identified as corresponding
clusters. The proportion of the organism's resources corresponding
to the technology cluster in the application to any biological organism
should be interpreted as the share of the organism 's resources that
allows performing the functions of the organism aimed at realizing
its energy and protective needs. For biological organisms, the term
``technology cluster'' can be renamed to ``transformation cluster''.

{}We will number the basic needs with an index $i$ (the values $i=1,2,3,4,5$
correspond to energy, transport, ecology, technology and information
needs, respectively). We will also denote by $C_{i}$ the value of
any extensive parameter (mass, time, etc.) that belongs to the cluster
(basic need) with the number $i$. The $C_{i}$ unit is the same as
that of its extensive parameter (mass, time, etc.). The $\bar{C}_{i}$
values (with a bar) are normalized and are measured in fractions (or
percentages) of the extensive parameter for the whole organism. Statistical
analysis in biological organisms allowed us to obtain the ideal distribution
in clusters. The corresponding reference (ideal) values $\bar{C}_{i}$
are determined in fractions of total resource $38\%$ $\left(0.38\pm0.06\right)$,
$27\%$ $\left(0.27\pm0.05\right)$, $16\%$ $\left(0.16\pm0.04\right)$,
$13\%$ $\left(0.13\pm0.02\right)$, $6\%$ $\left(0.06\pm0.01\right)$
for $i=1,2,3,4,5$ respectively, that is they are dimensionless quantities.
In the context of the study of a biological organism, the mass of
the organism is taken as an extensive parameter of $100\%$, and the
corresponding fractions of the organism mass are $0.38$, $0.27$,
$0.16$, $0.13$ and $0.06$ and correspond to the amount of resources
in the energy, transport, ecology, technology and information systems
of the organism. The sum of the five clusters expressed in units of
the system's resources, representing any extensive parameter (mass,
time, etc.), is a constant value ($100\%$) at a certain interval
of the organism life.

\begin{figure}[H]
{}\centering{}\includegraphics[scale=0.5]{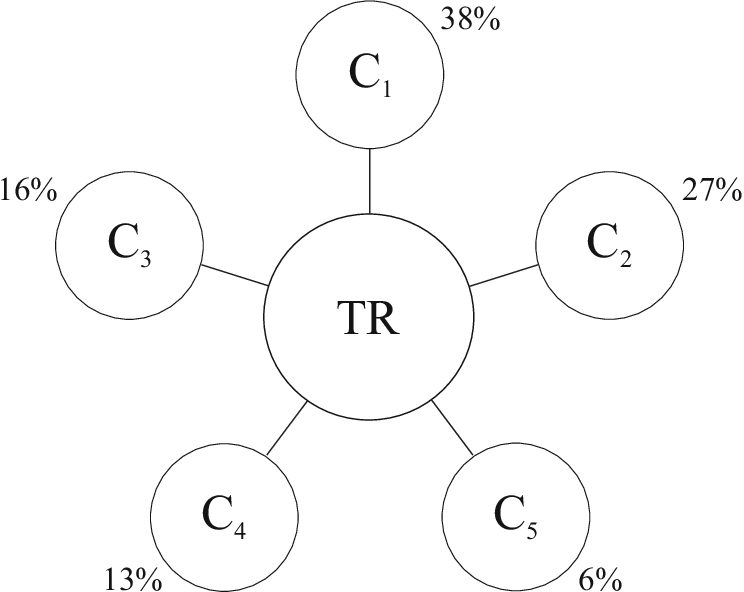}\caption{Fractal-cluster structuring of a complex system. \label{figure1} }
\end{figure}

\begin{figure}[H]
{}\centering{}\centering{}\includegraphics[scale=0.8]{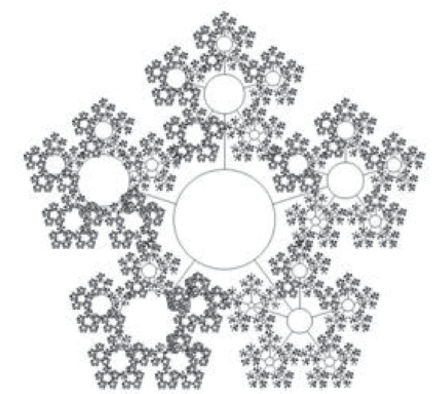}\caption{Schematic of the SOS's resource allocation for a hypothetical the
$6$-level fractal-cluster system. The figure is a tree graph with
the $6$ hierarchical levels and circled vertices, each circle being
a cluster of the corresponding hierarchy level. \label{figure2} }
\end{figure}

{}For a number of systems belonging to the class of organism (this
term will be explained in the next chapter in details), each cluster
can be a functioning subsystem in which basic needs can also be identified,
similar to those that can be identified in the organism. However,
the cluster of an organism cannot be considered as a separate organism,
since the cluster cannot function separately from the organism without
violating its integrity. We can divide each cluster into five subsystems
- subclusters, each of which has its own basic function, coinciding
with one of the basic needs of the entire system. For example, the
resource in the energy cluster can be shared for supporting its own
energy, transport, ecology, technology and information subclusters.
It can be expected that such a partition can be extended to subclusters
of the $n$-th level. However, for real systems it is very problematic
to identify subclusters of levels greater than $n=2$ because it requires
a separate study. For biological organisms, as a rule, two-level clustering
($n=2$) is sufficient because the number of hierarchy levels in real
systems is always finite and, as a rule, in real analysis, the number
of levels that can be identified does not exceed $n=3$. Namely, each
of the subclusters of a given level can be considered as the union
of the five higher-level subclusters. Such systems are called the
fractal-cluster (FC) systems (see \cite{Burd,Volov1,Volov2,Volov3}).
Thus, the space of the resource distribution for the same systems
has a hierarchical structure which can be described with a hierarchical
tree with a fixed number of branches $p=5$ \cite{VZ} (\fig{figure1}
and \fig{figure2}). \fig{figure1} and \fig{figure2} represent
an illustration of the resource distribution of an organism at one
(\fig{figure1}) and six levels (\fig{figure2}) of the hierarchy
of resource distribution in the $5^{n}$-dimensional FC space. A visual
illustration of a hypothetical 6-level fractal-cluster system is shown
in \fig{figure2}. Here we have a tree graph with $6$ hierarchy
levels and with vertices denoted by circles, each circle being a cluster
(or subcluster) of the corresponding hierarchy level. \fig{figure2}
corresponds to the idealized case and is a tree graph, with 6 hierarchy
levels and with vertices denoted by circles, each circle being a cluster
of the corresponding hierarchy level.

{}

{}

{}It should be emphasized that the fractal-cluster description of
an organism exclusively refers to the distribution of its resource
according to its basic needs.

{}The closest values of clusters to their ideal values were obtained
for young men under 25 years of age \cite{Burd}. These ideal (standard)
relations among the clusters were named the fractal-cluster relations
(FCR). In the fractal-cluster approach, the normal (effective) functioning
of an organism is understood as such functioning in which the mathematical
expectation of cluster values are close to the ideal (reference) values
of clusters. For the organism with normal functioning of any nature
(biological, social and economical) these ideal values have approximately
the same values. Specific mechanisms of the resource distribution
in fractal-cluster models have been presented in \cite{VZ}. A natural
question arises -- why are exactly the five basic needs involved
when defining such a concept as an organism in the context of our
approach? These needs (energy, transport, ecology (or security), technology
(or transformation) and information (or management) are basic, because
without their independent satisfaction (without outside help) the
organism cannot exist. From the point of view of the fractal-cluster
approach, an organism can be considered any self-organizing system
(SOS) in which the five basic needs can be identified that are satisfied
by the SOS, independently. However, one can note that the number five
also occurs in various areas of the natural sciences, and is also
a quantitative measure of the properties of the certain biological
systems. For example, it is known that thermodynamics of the stable
states has five thermodynamic potentials \cite{Presnov}. Man has
five sense organs, five brain rhythms, and in human anatomy the number
five is highlighted \cite{KE,Byk}. When considering flowers, when
the number of petals is a multiple of five, there is always an ovary,
otherwise there is no ovary. Other examples can be given, but of course
this is not any scientific justification for choosing the number five
as the number of clusters. In fact, the offered FC approach is qualitatively
different from the traditional approach for description of a complex
SOS. The offered approach \cite{Burd} was based on the fractal structure
basis -- the FC matrix -- allowing actively to describe the resource
distributions in the SOS. Unfortunately, both theory and mathematical
models on the basis of this approach have not been developed. That
is why in 1998 professor V. P. Burdakov offered his young colleague
(the author of this article) to start developing the fractal-cluster
theory.

{}The little fame of the offered approach in the broader scientific
community is explained by the fact that publications did only appear
in the Russian language. The aim of this article is on the one hand
getting biological scientists acquainted with the offered FC approach
and, on the other hand, exposition of new results obtained in the
development of this approach. The possibility of using the FC models
for the resource distribution analysis is based on the range researches
\cite{Burd,Volov1,Volov2,Volov3}, in which researches of the optimal
resource distribution have been presented. These methods are based
on the thermodynamic method in its informational interpretation. Now
one can show some conformations of validity of the FC structuring
for the SOS. These confirmations are the following: 1) the set of
experimental biological data analyzed by mathematical statistical
methods \cite{Burd}; 2) the theoretical value of the transport cluster
has been obtained by using the Limit energy theorem for gas flow systems
\cite{Volov4}.

{}The paper is organized as follows. In section \ref{Features} features
of identification of the term ``organism'' with the fractal-cluster
approach to research in biology and economy are presented. Section
\ref{Criteria} is devoted to the FC criteria of the fractal-cluster
theory. An analysis of the static and dynamic stability for the ``organism''
evolution are presented in section \ref{Analysis}. The determined
fractal-cluster theory is presented in section \ref{Det}. The stochastic
distributions in the FC theory are considered in section \ref{Prob}.
An applications of the fractal-cluster theory for the resource distribution
controlling in the socio-economical and biological systems are presented
in section \ref{App}. Discussion and prospects for further research
are analyzed in section \ref{Concl}.

\section[Features of identification of the term \textquotedblleft organism\textquotedblright{}]{Features of identification of the term \textquotedblleft organism\textquotedblright{}
in the fractal-cluster approach to biology and socio-economic researches}

{}\label{Features}

{}As noted above, in the fractal-cluster approach, the ``organism''
is understood as a complex SOS in which the five basic needs can be
identified: energy, transport, ecology, technology and information
needs. At the same time, the SOS (biology organisms, socio-economical
systems, systems of the ``machine-human'' class and etc.) must satisfy
its basic needs on its own. These needs are called clusters that are
named accordingly (energy cluster etc.). It should be noted that such
names of clusters have a historical imprint, and this is due to the
fact that the initial classification of resources was not for biological
organisms, but in technological systems. As a rule, we will use the
same names for clusters of biological systems. The technology cluster
can also be called a transformation cluster. Note also that there
are much more needs for the self-organizing systems, but all of them
can be ``packaged'' in the hierarchical fractal structure, with
the self-similar subclusters of $n$-levels. As a resource, any extensive
parameter of an organism (mass, time, volume, etc.) is used. From
the point of view of the fractal-cluster approach, any complex system
in which it is impossible to distinguish the five basic needs is not
an organism, and the announced approach cannot be applied to it. It
must be emphasized that resource distribution in the FC systems is
a nontrivial task which requires special skills and study. As an additional
explanation for the fractal-cluster distribution of resources, for
example, when choosing the organism mass as an extensive parameter,
a separate organism's structure can perform several functions, and
therefore belong to different clusters.

{}As an example of the most developed self-organizing system among
the biological systems can be considered the human organism. For this
system as the ``resource'' we can choose a share of time on a certain
time scale (for example, a period of 24 hours) needed for satisfying
the corresponding five basic needs. Bearing it in mind, a time share
spent on satisfaction of energy needs (sleep, food intake etc.) is
included in the energy cluster. The transport cluster consists of
a share of time spent on transfer. A share of time spent on rest,
wellness treatments etc. is included in the ecology cluster. Working
time represents the share of resources going to the technology cluster.
The information cluster includes the time spent on obtaining new knowledge
and skills. We have to emphasize that for healthy individual (18 --
64 years old) according to the National Sleep Foundation \cite{NSF}
sleep time is 7 -- 9 hours . Based on statistical data, it was shown
in \cite{Baran} that the meal time (without talking) should not be
less than 20 minutes, that is, with three meals a day, it is one hour.
Thus, the ratio of sleep and meal time to 24 hours is close to the
value of the ideal value of the energy cluster ($8/24\approx0.38$),
etc.

{}According to the fractal-cluster approach all movements should
take approximately 6.5 hours, rest takes 3.5 hours, technology or
activity is about 3 hours, effective receipt and processing of information
take about one and half hour. These assessments are correlated with
World Health Organization guidelines \cite{WHO}. If you use the organism's
weight (of a young person) as an extensive parameter, then, for example,
the transport cluster is a ratio of the weight of the legs and part
of the muscle apparatus to the weight of the entire body. The ratio
will be close on average to $0.27$, i.e. it is approximately equal
to the ideal value of the transport cluster \cite{Burd,Volov1,Volov2,Volov3}.
The technology cluster is a ratio of the weight of the arms and part
of the muscle apparatus to the weight of the entire body will be close
to the ideal value of the technology cluster 0.13 \cite{Burd,Volov1,Volov2,Volov3}.
This ratio of energy cluster including the respiratory, digestive,
fat and circulatory systems will be close on average to $0.38$ \cite{Burd,Volov1,Volov2,Volov3}.
The information cluster ratio which includes brain, sexual system
and nervous system will be close on average to $0.06$ \cite{Burd,Volov1,Volov2,Volov3}.
It should be noted that prolonged significant deviation of resource
values in clusters from their ideal (reference) values leads to disturbance
of the organism's work accompanying various pathologies \cite{Burd,Volov1,Volov2,Volov3}.
The necessary and sufficient condition of definition of the term ``organism''
on the basis of the fractal-cluster approach should be highlighted.
The necessary condition is the presence in a complex system of the
five above mentioned basic needs. A sufficient condition is the ability
of the ``organism'' itself to satisfy its basic needs. Thus, for
example, from the point of view of fractal-cluster theory a baby,
despite the presence of five basic needs, is not an organism, because
without the mother's help, the one cannot move, receive food, etc.
The definition of the organism in the fractal-cluster approach is
qualitatively different from the generally accepted one in biology
and biophysics. For example, classical factors of determining an organism
\cite{Campbell}: 1) cell structure, 2) reproduction, 3) growth and
development, 4) energy utilization, 5) response to the environment,
6) homeostasis, 7) evolutionary adaptation determines mainly external
characteristics and results of life activity of an organism interacting
with an external environment. The fractal-cluster approach in the
study of biological organisms determines the distribution of the resource
within an organism according to its basic needs (clusters) and it
does not determine the growth of an organism. However, the growth
factor of an organism is related to changes in the total resource
of the organism. To overcome this, the resource change curve is divided
into a number of sections in which the change in the organism's resource
can be neglected (quasi-homeostasis). In this case, at each such stage
of development of the organism, the criterial apparatus of the FC
theory is applicable. This approach is similar to the classical thermodynamic
method, when the nonequilibrium process is replaced by a series of
equilibrium processes when the initial and the final states of the
process coincide. The energy utilization factor 4) undoubtedly correlates
with the energy cluster of the fractal-cluster approach to organism
studies. Factors 5) -- 7) correlate with clusters of ecology (safety),
information (controlling) and technology (transformation). Environmental
impacts, such as lack of food, or internal causes (an organism's disease)
lead to a change in the distribution of resources in the corresponding
clusters. The organism, having received information about the imbalance
of resources (information cluster) carries out transformation of resources
(technology cluster), which in turn leads to a state of sustainable
equilibrium with an environment, i.e homeostasis. The homeostasis
factor correlates with oscillations of clusters and fractal-cluster
values near their standard (reference) values. A cellularly ordered
structure in a certain sense corresponds with the fractal-cluster
hierarchical structure of the organism's resource representation.
Thus, it is possible to summarize that, despite the difference from
the classical organism's definition, the introduced concept of the
organism at the fractal-cluster approach does not contradict the classical
representation of an organism in biology, on the one hand, and allows
us to develop a new toolkit for the study of organism development
on the other hand (\tabl{tab1}).

{}
\begin{table}[H]
{}\centering{}\caption{Basic characteristics of biological organisms when modeled by traditional
methods and tools of the fractal-cluster theory}
\label{tab1} \centering{}\vspace{1mm}
\begin{tabular}{|p{150pt}|p{80pt}|p{80pt}|}
\hline
 & {}Traditional description of biological organisms  & {}FC description of biological organisms\tabularnewline
\hline
{}Complex organisms, open system  & {}Yes  & {}Yes\tabularnewline
\hline
{}Biochemical, biophysical machines  & {}Yes  & {}Do not contradict\tabularnewline
\hline
{}Reacts as a whole to external influences  & {}Yes  & {}Yes\tabularnewline
\hline
{}Hierarchy of organisms  & {}Yes  & {}Yes, except for the level\tabularnewline
\hline
{}Reproduction  & {}Yes  & {}Does not contradict\tabularnewline
\hline
{}Transmission of hereditary information  & {}Yes  & {}No\tabularnewline
\hline
{}Variability  & {}Yes  & {}Does not contradict\tabularnewline
\hline
{}Individual development  & {}Yes  & {}Yes\tabularnewline
\hline
{}Evolution of development  & {}Yes  & {}Yes\tabularnewline
\hline
{}Rhythmicity  & {}Yes  & {}Yes\tabularnewline
\hline
{}The possibility of studying FC tools  & {}No  & {}Yes\tabularnewline
\hline
{}Identification of the organism as self-organizing system independently
satisfying basic needs  & {}No  & {}Yes\tabularnewline
\hline
\end{tabular}
\end{table}

\section{Criteria of the fractal-cluster theory}

{}\label{Criteria}

{}This section presents the FC criteria of resource distribution
for organisms based on the synthesis of the FCR and nonequilibrium
thermodynamics. This resource research of organisms is based on the
FC approach, and shows that the researched object in question is not
decomposed, but is a ``black box'', which corresponds to the principles
and methodology of thermodynamics. In the physical space the real
object has the resources $X_{i}$, necessary for its functioning,
and the results of its activities. While transferring the external
variables from the physical space to the $5^{n}$--dimensional FC
space, the decomposition and classification of information about the
object resources have been accomplished, i.e., the FC structurization
of the information about necessary resources for organisms (energy,
transport, ecology, technology and information support services).
As shown in \cite{Volov1,Volov2,Volov3} thermodynamic laws and theorems
make it possible to analyze the stability and the resourse allocation
efficiency of organisms, without additional empirical information.

{}The FC criteria of an organism are defined in a nontrivial way
with the construction of the fractal-cluster model. In the FC space,
the cluster values $\left\{ C_{i}\right\} $ and $n$ level subclusters
values $\left\{ C_{i_{1}i_{2}\cdots i_{n}}\right\} $ for any $n$
are a positive numbers. As defined above unit is the same as the extensive
parameter (mass, time, etc.), see Sec. \ref{Introduction}. Further
it is convenient to enter the normalized values of the resources in
the clusters:

{}
\[
\overline{C}_{i_{1}i_{2}\cdots i_{n}}=\dfrac{C_{i_{1}i_{2}\cdots i_{n}}}{C_{\Sigma}},\;C_{\Sigma}=\sum_{i=1}^{5}C_{i}.
\]
The values $\overline{C}_{i_{1}i_{2}\cdots i_{n}}$ form a $n$-dimensional
matrix, which we will call the resource allocation matrix or the fractal-cluster
matrix (FCM). The permissible clusters value scope is defined as follows

{}
\begin{equation}
0<\overline{C}_{i_{1}i_{2}\cdots i_{n}}<a_{i_{1}i_{2}\cdots i_{n}}\;\mathrm{for\:all\:}n\geq1,\label{C_restr}
\end{equation}
where $0<a_{i_{1}i_{2}\cdots i_{n}}<1$ and conservation law holds:

{}
\begin{equation}
\sum_{i_{1}=1}^{5}\sum_{i_{2}=1}^{5}\ldots\sum_{i_{n}=1}^{5}\overline{C}_{i_{1}i_{2}\cdots i_{n}}=1.\label{conserv}
\end{equation}
Equations (\ref{C_restr}) determine the range of changes in the value
of clusters and subclusters of the organism \cite{Burd,Volov1,Volov2,Volov3,Burd_Th}:
their positivity ($\overline{C}_{i_{1}i_{2}\cdots i_{n}}>0$), and
their limitedness ($\overline{C}_{i_{1}i_{2}\cdots i_{n}}<a_{i_{1}i_{2}\cdots i_{n}}$,
where $a_{i_{1}i_{2}\cdots i_{n}}<1$) for all $n$. The equation
(\ref{conserv}) determines the law of conservation of the organism's
resource at a given stage of the organism's development. The Section
\ref{Prob} presents an algorithm for determining the coefficients
$a_{i_{1}i_{2}\cdots i_{n}}$, based on possible values of cluster
combinations and the conservation law (\ref{conserv}).

\noindent {}The FC criteria are based on:

{}1) the axiom of the FCR universality (the organism's five-cluster
structuring of the resource needs);

{}2) the assumption that values of resources in clusters cannot take
zero values.

{}Initially, the FC approach which has been proposed in \cite{Burd}
was based on intuitive concepts and analogies and then on hard concepts
\cite{Volov1}. In connection with the above, it is logical to formulate
a criterion for the efficiency of the FC matrix on the fundamental
principles and methods of thermodynamics of stable states.

{}One of the most important quantities in the FC theory is the FC
entropy, which is an energy function of resource distribution. The
formal definition of this function is as follows. To determine the
FC entropy $S$ for a general $n$-level system, we will impose the
following requirements on it \cite{ME}:

{}1) additivity;

{}2) positive definiteness;

{}3) finiteness.

{}In addition, we will require that:

{}4) the FC entropy depends only on the resource values in the energy
cluster as well as the resource values in the higher level energy
subclusters of all non-energy clusters.

{}Requirement 1) is satisfied if $S$ is a linear homogeneous function
of resource values in the energy cluster and in the higher level energy
subclusters of non-energy clusters. Then requirements 2) and 3) are
satisfied if a simple sum of resource values is chosen as such a linear
function. Given requirement 4), this leads to the following expression
for $S$:
\[
S=C_{1}+\sum_{i_{1}=2}^{5}\left(\sum_{i_{2}=1}^{5}\ldots\sum_{i_{n-1}=1}^{5}C_{i_{1}i_{2}\ldots i_{n-1}1}\right).
\]
Because $S$ is finite, it is convenient to work with the normalized
entropy $H$ instead, which is $S$ divided by the value of the total
resource $C_{\Sigma}$:

{}
\begin{equation}
H=\dfrac{S}{C_{\Sigma}}=\overline{C}_{1}+\sum_{i_{1}=2}^{5}\left(\sum_{i_{2}=1}^{5}\ldots\sum_{i_{n-1}=1}^{5}\overline{C}_{i_{1}i_{2}\ldots i_{n-1}1}\right),\label{H_n}
\end{equation}
We will call this value the normalized $n$-level fractal-cluster
entropy and denote also by $H_{n}$. The proposed expression for FC
entropy has the following meaning: it is the proportion of the organism's
resources used to satisfy all its energy needs in all clusters.

{}It is necessary to clarify that the proposed mathematical measure
is a fractal-cluster entropy differs significantly from the usual
forms of entropy recording, which have a logarithmic deterministic
or probabilistic form. However, the basic characteristics of fractal
cluster entropy completely correspond to the properties of the construction
of this function \cite{ME}. An example of the use in physics of a
non-logarithmic form of entropy is the entropy of black holes introduced
by J.D. Bekenstein, which is proportional to the area of their event
horizon \cite{Bek}.

{}Let us consider the matrix of FC ideal states (two-dimensional
case $n=2$, \tabl{tab2}). The first column numbers the rows of
the matrix. In this case, the line numbers $i=1,2,3,4,5$ correspond
to the numbers of the clusters (1 -- energy, 2 -- transport, 3 --
ecology, 4 -- technology, 5 -- information). The second column of
the $i$-th row contains the ideal values of these clusters. Each
cluster is divided into subclusters, and the 3, 4, 5, 6, and 7 columns
of the $i$-th row contain the ideal values of the subclusters of
the $i$-th cluster. The sum of the values of the subclusters of the
first row and the first column of the resources of the ideal matrix
give quantitative information about the total share of the organism's
energy resources, which is equal to $0.615$. This number is very
close to the so-called ``Golden Ratio'' $0.618$, known from numerous
publications as the basis of beauty and harmony of both natural and
anthropogenic phenomena \cite{Stakhov1,Stakhov2,Stakhov0,GGZ}.

{}
\begin{table}[H]
{}\centering{}\centering{}\centering{}\caption{Table of the ideal cluster values.}
\vspace{1mm}
\begin{tabular}{|c|c|c|c|c|c|c|}
\hline
{}$j$  & {}$\overline{C}_{j}^{ideal}$  & {}$\overline{C}_{1j}^{ideal}$  & {}$\overline{C}_{2j}^{ideal}$  & {}$\overline{C}_{3j}^{ideal}$  & {}$\overline{C}_{4j}^{ideal}$  & {}$\overline{C}_{5j}^{ideal}$\tabularnewline
\hline
{}\, 1 \,  & {}\, 0.38 \,  & {}\, 0.144\,  & {}\,0.1026\,  & {}\,0.0608\,  & {}\, 0.0494 \,  & {}\,0.0228\,\tabularnewline
\hline
{}\, 2 \,  & {}\,0.27\,  & {}\, 0.1026\,  & {}\,0.0729\,  & {}\, 0.0432 \,  & {}\,0.0351 \,  & {}\, 0.0162\,\tabularnewline
\hline
{}\, 3 \,  & {}\,0.16\,  & {}\,0.0608\,  & {}\,0.0432\,  & {}\, 0.0256\,  & {}\, 0.0208\,  & {}\, 0.096\,\tabularnewline
\hline
{}\, 4 \,  & {}\, 0.13 \,  & {}\,0.0494\,  & {}\, 0.0351 \,  & {}\, 0.208\,  & {}\,0.0169 \,  & {}0.078\,\tabularnewline
\hline
{}\, 5 \,  & {}\, 0.06\,  & {}\,0.0228\,  & {}\,0.0169\,  & {}\,0.096\,  & {}\,0.078\,  & {}\,0.0036\,\tabularnewline
\hline
\end{tabular}

\label{tab2}
\end{table}

{}The ideal FCM is symmetric for the reference values. The elements
of the matrix are $\overline{C}_{ij}^{ideal}=\overline{C}_{i}^{ideal}\overline{C}_{j}^{ideal}$.
At the same time $\sum_{j=1}^{5}\overline{C}_{ij}^{ideal}=\overline{C}_{i}^{ideal}.$
The total share of the organism's energy resources being the major
determinant of the organism's functioning effectiveness is determind
by the following formula:

{}
\[
\overline{C}_{\Sigma}^{energy}=\sum_{j=1}^{5}\overline{C}_{1j}^{ideal}+\sum_{i=2}^{5}\overline{C}_{i1}^{ideal}=\overline{C}_{1}^{ideal}+\overline{C}_{1}^{ideal}\sum_{i=2}^{5}\overline{C}_{i}^{ideal}=
\]

{}
\begin{equation}
=\overline{C}_{1}^{ideal}+\overline{C}_{1}^{ideal}\left(1-\overline{C}_{1}^{ideal}\right)=2\overline{C}_{1}^{ideal}-\left(\overline{C}_{1}^{ideal}\right)^{2}\approx0.615\equiv H_{0}.\label{H_0}
\end{equation}
The expression (\ref{H_0}) is nothing but the entropy of a 2-level
FCM with an ideal distribution of resources.This formula presents
resource's distribution of an organism obtained in \cite{Burd,Volov3}.
For the nonideal distribution of the organism resources the fractal-cluster
entropy for the two-dimentional FCM has the following form:

{}
\begin{equation}
H_{2}\equiv\sum_{j=1}^{5}\overline{C}_{1j}+\sum_{i=2}^{5}\overline{C}_{i1}=\overline{C}_{1}+\overline{C}_{1}\sum_{i=2}^{5}\overline{C}_{i}.\label{H_2}
\end{equation}

{}The relationship among the FCM elements for the ideal and nonideal
cases and the fractal-cluster entropy $H_{2}$ noted above allows
us to find a solution of the FCM for the purpose of optimal evolution
$\overline{C}_{1j}=\overline{C}_{1j}\left(t\right)$ ($j=1,\ldots,5$)
and $\overline{C}_{i1}=\overline{C}_{i1}\left(t\right)$ ($i=2,\ldots,5$)
from the nonideal state of the organism (a nonideal FCM) to the perfect
condition -- (an ideal FCM), so that, the sum of the FCM elements
of the first column and the first row (\ref{H_0}) goes into their
ideal value, that is, the ``Golden Ratio'' entropy value is achieved:
$H_{2}\left(\left\{ \overline{C}_{ij}\left(t\right)\right\} \right)\rightarrow H_{0}$.

{}The above proposed criterion of the FC entropy $H$ (\ref{H_n})
can be attributed to the static criteria. In addition, the criteria
of full effectiveness $\eta^{\Sigma}$, proposed in \cite{Burd},
can be treated as the static criteria of organism's full effectiveness.
The full effectiveness $\eta^{\Sigma}$ of an organism defined as
the minimum ratio $\dfrac{\overline{C}_{i}}{\overline{C}_{i}^{ideal}}$
so named Libih's barrel \cite{Burd}:

{}
\begin{equation}
\eta^{\Sigma}=\min_{i}\left(\dfrac{\overline{C}_{i}}{\overline{C}_{i}^{ideal}}\right).\label{eta}
\end{equation}
Equation (\ref{eta}) is a criterion for the efficiency of the life
of an organism, introduced in \cite{Burd}.

{}To estimate the maximum work done by the system, we used the thermodynamic
Helmholtz potential of the system $F$ \cite{Volov1}, defined as
follows

{}
\[
F=U-TS,
\]
where $U$ is the internal energy of the system, $T$ is the temperature,
and $S$ is the thermodynamic entropy. As it is known this potential
is called the energy. This potential characterizes the maximum possible
work that the system can perform.

{}An analogue of the dimensionless Helmholtz potential in the fractal-cluster
models \cite{Volov1,Volov3} is the free fractal-cluster energy of
the organism defined by the following expression
\begin{equation}
F=\left|\overline{C}_{1}-H\right|.\label{F}
\end{equation}
But these criteria $\left(H,\:\eta^{\Sigma},\:F\right)$ are not sensitive
enough that is, with small changes in clusters, small changes in the
criteria take place. This fact does not allow predicting in advance
the crisis tendencies of the functioning of the organism.

{}To determine a highly sensitive criterion of the organism's resource
distribution, Hausdorff's approach was used. In contrast to the purely
fractal structures, the FC $n$-dimensional matrix substantially differs
from the geometrical fractal structures, as the quantitative distribution
in subclusters of any level may differ from the ideal distribution
and, thus, the organism quality changes. Therefore, the following
algorithm to determine the highly sensitive criterion of the FC effectiveness
has been proposed in \cite{Volov3}. The FC criterion of the resource
distribution effectiveness is determined with the formula

{}
\begin{equation}
D=\frac{\log\overset{5}{\underset{i_{1}=1}{\sum}}\overset{5}{\underset{i_{2}=1}{\sum}}\ldots\overset{5}{\underset{i_{n}=1}{\sum}}\delta_{i_{1}i_{2}\cdots i_{n}}^{\star}}{\log N},\label{D_crit}
\end{equation}
where $N$ is a total number of clusters and subclusters and values
$\delta_{i_{1}i_{2}\cdots i_{n}}^{\star}$are calculated with the
relations

{}
\begin{equation}
\delta_{i_{1}i_{2}\cdots i_{n}}^{\star}=1-\dfrac{\left|\overline{C}_{i_{1}i_{2}\cdots i_{n}}-\overline{C}_{i_{1}i_{2}\cdots i_{n}}^{ideal}\right|}{\overline{C}_{i_{1}i_{2}\cdots i_{n}}^{ideal}},\label{delta_star}
\end{equation}
where $\overline{C}_{i_{1}i_{2}\cdots i_{n}}^{ideal}=\overline{C}_{i_{1}}^{ideal}\overline{C}_{i_{2}}^{ideal}\cdots\overline{C}_{i_{n}}^{ideal}$.

{}There were the following reasons for definition of the $D$-criterion
presented in \cite{Volov3}: since the resource space is fractal,
then, by analogy with fractal geometry, where the dimension of the
space is determined by the $D$ -- Hausdorff dimension criterion,
it was logical to introduce the corresponding criterion, where the
resource metric was introduced instead of the geometric metric (deviation
of the values of resources in the hierarchical structure of the fractal-cluster
matrix of the organism from their reference values). The $D$-criterion
is qualitatively different from the Hausdorff dimension criterion:
The $D$-criterion, in contrast to the Hausdorff dimension $D$, can
take not only integer and fractional values, but also negative values.
In addition, the $D$-criterion obeys the resource conservation law
(\ref{conserv}). An approbation of the $D$-criterion showed its
high sensitivity compared to the FC entropy H and the $D$-criterion.
The efficiency of resource allocation (the $D$-criterion) of an organism
is understood as a deterministic measure of the deviation of the values
of its clusters and subclusters from their reference values.

{}A mixed FC criterion of organisms was presented in \cite{Volov3},
as follows

\begin{figure}[H]
{}\centering{}\centering{}\includegraphics[scale=0.7]{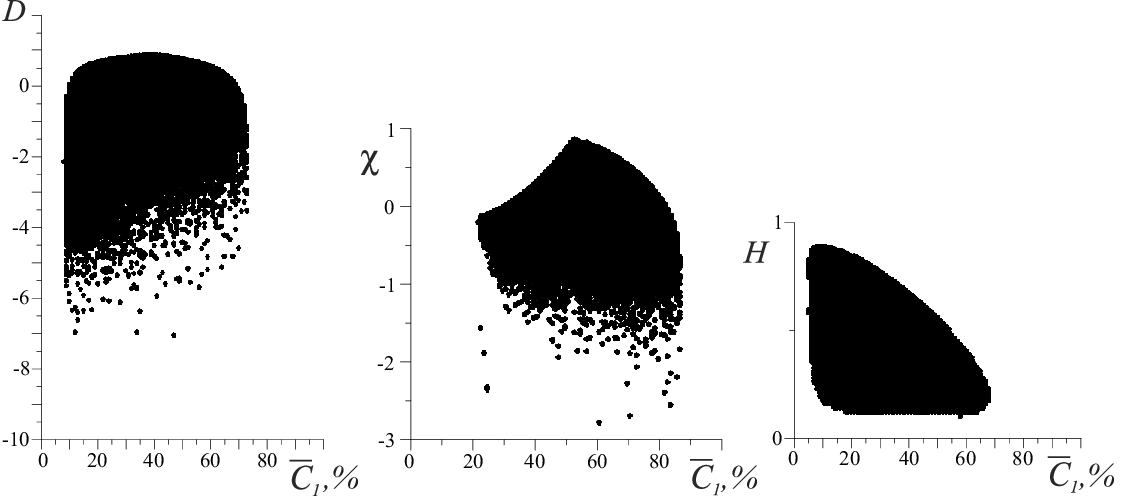}\caption{The range of possible values of the \emph{$D$}, $\chi$ and FC entropy
$H$ for cluster $\overline{C}_{1}$. The range possiable states of
organisms corresponds to equation (\ref{C_restr}). The vertical axes
represent values of the $D$-criterion, the mixed $\chi$-criterion,
and the FC entropy $H$, respectively. The horizontal axes represent
values of the energy cluster $\overline{C}_{1}$ as a percentage.
\label{figure3} }
\end{figure}

\begin{figure}[H]
{}\centering{}\centering{}\includegraphics[scale=0.5]{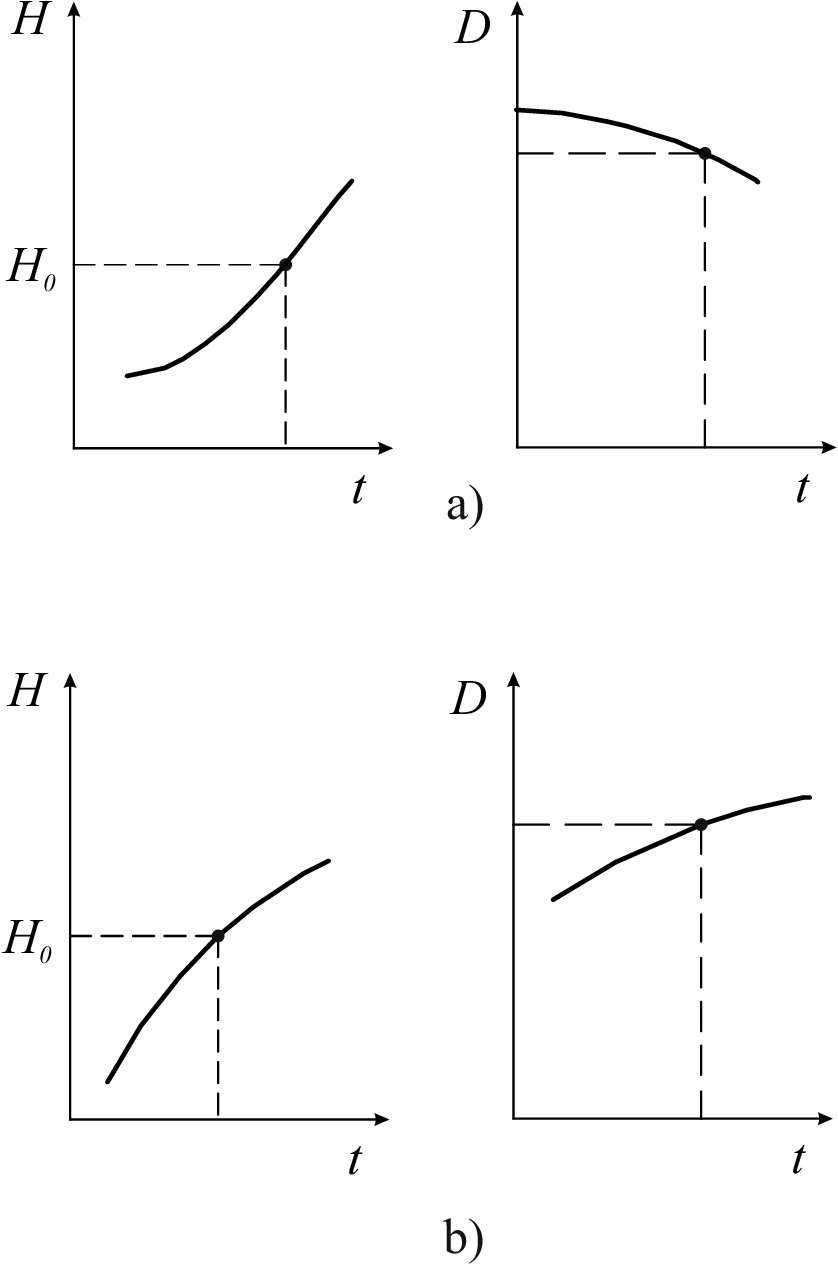}\caption{Changing of the FC criteria in time: a negative transformation of
the SOS (a); a positive state and transformation of the SOS (b). The
vertical axes represent values of the FC entropy $H$ and $D$-criterion,
respectively. The horizontal axis represents the time. \label{figure4} }
\end{figure}

{}
\begin{equation}
\chi=\frac{H\cdot D\cdot\eta^{\Sigma}}{H_{0}\cdot D^{\max}}.\label{hi}
\end{equation}
Unlike the entropy $H$, full effectiveness $\eta^{\Sigma}$, the
free energy $F$, $D$-criterion and \emph{{}$\chi$}{} are a very
sensitive indicator of varying the FCM values (\fig{figure3}). The
area outside the boundary (hatched area) is a nonfunctional state
of the organism. \fig{figure3}a shows that in the sector of negative
values of the \emph{{}$D$}{} and $\chi$ criteria, at the boundary
of the system, we have the destruction of the space continuity where
the FCM parameters can vary. This phenomenon can be interpreted as
the boundary where irreversible damage of the organism functioning
appears. This effect of destruction of the FC structure continuity
can be interpreted as a phenomenon of the FC percolation (\fig{figure3}).
\fig{figure4} shows an example of the sufficient and necessary conditions
of the resource distribution effectiveness. From these figures it
is clear that the point $H_{0}$ on the entropy trend (on the left
\fig{figure4} a) is the unstable state of the FC resource distribution
of the organism and the point $H_{0}$ on the entropy trend (on the
left \fig{figure4} b) is the stable state of the one.

{}

{}

From the point of view of the fractal-cluster approach, a necessary
condition for the effective functioning of self-organizing systems
of the ``organism'' class on a local time interval $\Delta T$ ($\Delta T\ll T$,
where $T$ is the total time of functioning of the organism) is the
proximity of the value of the fractal-cluster entropy H to the value
of the entropy of the ``Golden Ratio'' ($\left|H-H_{0}\right|\rightarrow0$).
A sufficient condition for the sustainable effective functioning of
the self-organizing systems in the local time interval has two components.
The first component is, according to the minimum entropy production
theorem \cite{PN}, the fulfillment of the entropy criterion ($\dfrac{d^{2}H}{dt^{2}}<0$),
i.e. the trend of entropy should be convex. This fact is confirmed
by an increase in criterion $D$ when this criterion is fulfilled
(\fig{figure4}b) and a decrease in criterion $D$ when it is not
fulfilled (\fig{figure4}a). The second component of a sufficient
condition is the fulfillment of the growth condition of the $D$-criterion
($D\rightarrow D_{\max}$).

\section{Analysis of the static and dynamic stability for organism's evolution}

{}\label{Analysis}

{}The stability analysis of the biological species' evolution can
be investigated on the basis of the fractal-cluster entropy. For this
purpose, it is necessary to use the apparatus of I. Prigogine's thermodynamics
of structure \cite{PN,GP} and nonlinear nonequilibrium fluctuation-dissipation
thermodynamics \cite{Strat} -- the minimum entropy production theorem
for the states close to the equilibrium state. For the states far
from equilibrium, a quadratic oscillating form -- a criterion for
the production of excess entropy -- is used.

{}We consider clusters $\left\{ \overline{C}_{i}\right\} $ and subclusters
$\left\{ \overline{C}_{ij}\right\} $, components of the organism,
as random internal parameters $C_{i}\left(t\right)$, $C_{ij}\left(t\right)$
changing in a fluctuational manner. If the organism is isolated, the
2-level FC entropy $H_{2}=H_{2}\left(\left\{ \overline{C}_{ij}\right\} \right)$
does not decrease with time. However, as shown in \cite{Strat}, micro
disturbance of the Second Law of thermodynamics for organisms cannot
exceed the value of $k$ ($k$ is Boltzmann's constant). This fact
allows us to obtain an assesment of the fluctuational component for
the fractal-cluster entropy $H_{2}\left(\left\{ \overline{C}_{ij}\right\} \right)$:

{}
\begin{equation}
\sqrt{\left(\delta H_{2}\left(\overline{C}_{ij}\left(t\right)\right)\right)^{2}}<k.\label{sqrt_lt_k}
\end{equation}

{}Let us divide the evolution time into $N$ $\left(N\gg1\right)$
identical intervals $\varDelta t$. By doing subcluster values averaging
on each intervals

{}
\begin{equation}
\left\langle \overline{C}_{ij}\right\rangle \left(t\right)\equiv\dfrac{1}{\varDelta t}\intop_{0}^{\varDelta t}\overline{C}_{ij}\left(t+t'\right)dt'\label{C_aver}
\end{equation}
we go to a new time variable $\tau$:
\[
\tau\left(t\right)=\left[\dfrac{t}{\varDelta t}\right]\varDelta t,
\]
where the symbol $\left[\cdots\right]$ denotes integer part. Formulas
(\ref{sqrt_lt_k}) and (\ref{C_aver}) refer to the estimates of entropy
fluctuations, according to \cite{Strat} and the averaging of the
cluster values over a certain time interval. The FC entropy $H_{2}=H_{2}\left(\tau\left(t\right)\right)$
becomes a step function of $t$ and in the case of asymmetric FCMs
has the following view:

{}
\begin{equation}
H_{2}\left(\tau\left(t\right)\right)=\left\langle \overline{C}_{1}\right\rangle +\sum_{j=2}^{5}\left\langle \overline{C}_{1j}\right\rangle .\label{H2_aver}
\end{equation}
The formula (\ref{H2_aver}) refers to the determination of the relationships
between clusters and subclusters of a two-dimensional symmetric FCM.

{}It is assumed that the time interval $\varDelta t$ is much less
than evolution time $T$ from the initial state$\left\{ \overline{C}_{ij}^{0}\right\} $
to the final (ideal) organism's state $\left\{ \overline{C}_{ij}\right\} ^{fin\left(ideal\right)}$

{}
\begin{equation}
\varDelta t\ll T,\;\left\langle \overline{C}_{ij}^{0}\right\rangle \:\stackrel{T}{\longrightarrow}\:\left\langle \overline{C}_{ij}\right\rangle ^{fin\left(ideal\right)}.\label{delta_t}
\end{equation}
In the symmetric case, subclusters $\left\langle \overline{C}_{ij}\right\rangle $
are determined with the relations

{}
\begin{equation}
\left\langle \overline{C}_{ij}\right\rangle =\left\langle \overline{C}_{ji}\right\rangle \;\textrm{and}\;\left\langle \overline{C}_{ij}\right\rangle =\left\langle \overline{C}_{i}\right\rangle \cdot\left\langle \overline{C}_{j}\right\rangle ,\;\textrm{i.e}.\;\left\langle \overline{C}_{ii}\right\rangle =\left\langle \overline{C}_{i}\right\rangle ^{2}.\label{C_aver_2}
\end{equation}
The FC entropy in this case is

{}
\begin{equation}
H_{2}=2\left\langle \overline{C}_{1}\right\rangle -\left\langle \overline{C}_{1}\right\rangle ^{2}.\label{H2_2}
\end{equation}

{}In accordance with the criterion of thermodynamic stability \cite{GP}
the second differential of the FC entropy $H_{2}$ for the symmetric
case is defined the following way

{}
\begin{equation}
\delta^{2}H_{2}=\frac{\partial^{2}H_{2}}{\partial\left\langle \overline{C}_{1}\right\rangle ^{2}}\left(\delta\left\langle \overline{C}_{1}\right\rangle \right)^{2}=-2\left(\delta\left\langle \overline{C}_{1}\right\rangle \right)^{2}\leq0.\label{delta_H_sym}
\end{equation}
\\
 {} This formula is a special case of formula (\ref{delta_H_asym})
for a symmetric FCM (see below).

{}Thus, for the states close to thermodynamic equilibrium for the
symmetric FCM, the second differential of the entropy $\delta^{2}H_{2}$
is negative, i.e., the organism is stable. The loss of stability for
the symmetric FCM is realized only when $\delta\left\langle \overline{C}_{1}\right\rangle =0$,
that is, in the complete absence of oscillations of the energy cluster
$\left\langle \overline{C}_{1}\right\rangle $.

{}In all other cases of the symmetrical FCM in the states close to
the branch of the thermodynamic equilibrium the stability criterion
is satisfied: $\delta^{2}H_{2}<0$.

{}In the case of the asymmetrical FCM the second differential of
the FC entropy has the form

{}
\[
\delta^{2}H_{2}\left(\left\langle \overline{C}_{1}\right\rangle ,\left\langle \overline{C}_{21}\right\rangle ,\left\langle \overline{C}_{31}\right\rangle ,\left\langle \overline{C}_{41}\right\rangle ,\left\langle \overline{C}_{51}\right\rangle \right)=
\]
\[
=\frac{\partial^{2}H}{\partial\left\langle \overline{C}_{1}\right\rangle ^{2}}\left(\delta\overline{C}_{1}\right)^{2}+\sum_{j=2}^{5}\frac{\partial^{2}H_{2}}{\partial\overline{C}_{j1}^{2}}\left(\delta\left\langle \overline{C}_{j1}\right\rangle \right)^{2}+
\]
\[
+2\frac{\partial}{\partial\left\langle \overline{C}_{1}\right\rangle }\sum_{j=2}^{5}\left(\frac{\partial H_{2}}{\partial\left\langle \overline{C}_{j1}\right\rangle }\cdot\delta\left\langle \overline{C}_{j1}\right\rangle \right)\left(\delta\left\langle \overline{C}_{1}\right\rangle \right)+
\]
\begin{equation}
+2\sum_{j=2}^{5}\sum_{j>i}^{5}\frac{\partial^{2}H_{2}}{\partial\left\langle \overline{C}_{i1}\right\rangle \partial\left\langle \overline{C}_{j1}\right\rangle }\delta\left\langle \overline{C}_{i1}\right\rangle \delta\left\langle \overline{C}_{j1}\right\rangle .\label{delta_H_asym}
\end{equation}

{}In the case of independence of energy cluster $\left\langle \overline{C}_{1}\right\rangle $
and energy subclusters $\left\langle \overline{C}_{21}\right\rangle ,$
$\left\langle \overline{C}_{31}\right\rangle ,$ $\left\langle \overline{C}_{41}\right\rangle $
and $\left\langle \overline{C}_{51}\right\rangle $, the second differential
of the FC entropy $\delta^{2}H_{2}$ is determined as follows
\begin{equation}
\delta^{2}H_{2}=0,\label{delta_2_H}
\end{equation}
that is, even in the presence of fluctuations, neutral stability of
the organism's evolution occurs. In the case of linear dependence
of the energy cluster $\left\langle \overline{C}_{1}\right\rangle $
and energy subclusters $\left\langle \overline{C}_{21}\right\rangle ,$
$\left\langle \overline{C}_{31}\right\rangle ,$ $\left\langle \overline{C}_{41}\right\rangle ,$
$\left\langle \overline{C}_{51}\right\rangle $ neutral stability
also occurs. In the case of nonlinear dependence of subclusters $\left\{ \left\langle \overline{C}_{ij}\right\rangle \right\} \:\left(i>1\right)$
from the energy cluster, there may appear both stable and unstable
regimes of the FCM evolution, i.e.
\begin{equation}
\delta^{2}H_{2}\left\{ \begin{array}{c}
<0-\textrm{stable\:\ regime},\\
=0-\textrm{neutral\:\ stability,}\\
>0-\textrm{unstable\:\ regime.}
\end{array}\right.\label{delta_2_H_stable}
\end{equation}
The system (\ref{delta_2_H_stable}) defines conditions of stability
according to Prigogine's theory \cite{GP}.

{}The analysis of the organism's stability given above which is based
on generalized thermodynamics of irreversible processes (Prigogine
\cite{PN,GP}) and the proposed FC theory applies to the states, close
to thermodynamic equilibrium branch, that is, to linear thermodynamics
of irreversible processes. The criterion of stability for organisms,
relevant to the concept of ``dissipative structures'' given by Prigogine,
is a quadratic oscillating form, called the excess entropy production
\cite{GP}. For the stable dissipative structures, the excess entropy
production is a positively definite value
\begin{equation}
P\left[\delta^{2}H\right]\equiv\dfrac{1}{2}\dfrac{d}{dt}\left(\delta^{2}H\right)>0,\label{P_delta_2_H}
\end{equation}
where $H=H\left(t\right)$ is the entropy of a general organism. The
formula (\ref{P_delta_2_H}) determines the expression of the criterion
of organism stability (the excess entropy production) located far
from the state of equilibrium \cite{GP}.

{}As it has been noted in \cite{PN}, in general case the sign of
excess entropy production cannot be determined unequivocally. To determine
$P\left[\delta^{2}H\right]$ sign the use of phenomenological laws
is required.

{}For the FC description of organism's resource structure, located
far from equilibrium, the following expression for the quadratic oscillating
form has been obtained, i.e., for the production of excess entropy
(or quasi-entropy) for the symmetric case of FCM $\left(\overline{C}_{ij}=\overline{C}_{ji}\right)$

{}
\begin{equation}
P\left[\delta^{2}H_{2}\right]=-\dfrac{\left[\delta\left\langle \overline{C}_{1}\right\rangle \left(t+\varDelta t\right)\right]^{2}-\left[\delta\left\langle \overline{C}_{1}\right\rangle \left(t\right)\right]^{2}}{\varDelta t}.\label{P_delta_2_H_2}
\end{equation}
The sign of the oscillating form $P\left[\delta^{2}H_{2}\right]$
is determined from the sign of the right-hand side of the equation
(\ref{P_delta_2_H_2}).

According to I. Prigogine's theorem \cite{PN}, the criterion of stability
far from equilibrium at the full time of self-organizing systems operation
is the fulfillment of the criterion:

\[
\dfrac{d\left(\delta^{2}H\right)}{dt}>0,
\]
where $H=2\overline{C}_{1}-\overline{C}_{1}^{2}$, $\delta H=\dfrac{\partial H}{\partial\overline{C}_{1}}\delta\overline{C}_{1}=2\delta\overline{C}_{1}\left(1-\overline{C}_{1}\right),$
$\delta^{2}H=-2\left(\delta\overline{C}_{1}\right)^{2},$ hence $-2\dfrac{d}{dt}\left(\delta\overline{C}_{1}\right)^{2}>0$.
The relative level of entropy oscillations is determined as follows

\[
\dfrac{\delta H}{H}=\dfrac{2\delta\overline{C}_{1}\left(1-\overline{C}_{1}\right)}{2\overline{C}_{1}-\overline{C}_{1}^{2}}.
\]
Thus, for a symmetric fractal-cluster matrix of the self-organizing
system which places far from the equilibrium, the stability condition
has the following form:

\begin{equation}
\dfrac{d}{dt}\left(\delta\overline{C}_{1}\right)^{2}<0\;-\;\mathrm{is\:the\:steady\:state,}\label{SS}
\end{equation}

\begin{equation}
\dfrac{d}{dt}\left(\delta\overline{C}_{1}\right)^{2}>0\;-\;\mathrm{is\:the\:unsteady\:state,}\label{USS}
\end{equation}

\begin{equation}
\dfrac{d}{dt}\left(\delta\overline{C}_{1}\right)^{2}=0\;-\;\mathrm{the\:neutral\:stability\:mode.}\label{NSM}
\end{equation}

In expressions (\ref{SS}) -- (\ref{NSM}), the value is a Lyapunov's
function for energy cluster oscillations, which is an sign alternating
quantity. Condition (\ref{SS}) means that the decrease in energy
cluster oscillations over time is a criterion for the longitudinal
stability of the functioning of the self-organizing system, which
is far from equilibrium. This fact fully corresponds to Lyapunov's
theory of stability.

For self-organizing systems of the ``organism'' class, an additional
restriction is imposed on the specified local-temporal and longitudinal
criteria for the stability of functioning, which determines the permissible
range of oscillations in the entropy trend. \tabl{tabnew} presents
scenarios for the longitudinal functioning of the self-organizing
system.

{}
\begin{table}[H]
{}\caption{Dynamic criteria for the sustainability of the transformation of resource
allocation in the self-organizing system over the entire time lag
of its operation.}
\centering{}%
\begin{tabular}{|>{\raggedright}m{4cm}|>{\raggedright}m{4cm}|>{\raggedright}m{4cm}|}
\hline
{}Transformation mode  & {}Mathematical description of the transformation process  & {}Notes\tabularnewline
\hline
{}Stable mode  & {}$\dfrac{d}{dt}\left(\delta\overline{C}_{1}\right)^{2}<0$,

{}$\left|\dfrac{\delta H}{H}\right|\ll1$  & {}$\delta\overline{C}_{1}$ sign alternating oscillation of an energy
cluster. As a rule, it is accepted $\left|\dfrac{\delta H}{H}\right|<0.1$\tabularnewline
\hline
{}Unstable mode  & {}$\dfrac{d}{dt}\left(\delta\overline{C}_{1}\right)^{2}>0$  & {} $\left|\dfrac{\delta H}{H}\right|\sim1$ -- the abnormal instability;

{}$\left|\dfrac{\delta H}{H}\right|\ll1$ -- the functional instability \tabularnewline
\hline
{}Neutral mode  & {}$\dfrac{d}{dt}\left(\delta\overline{C}_{1}\right)^{2}=0$  & {}Linear change of the fractal-cluster entropy $H$ over time;

{}$\left|\dfrac{\delta H}{H}\right|\ll1$ -- the normal neutral
mode;

{}$\left|\dfrac{\delta H}{H}\right|\sim1$ -- the abnormal neutral
mode of the self-organizing system operation.\tabularnewline
\hline
\end{tabular}

\label{tabnew}
\end{table}

Thus, as follows from the \tabl{tabnew} , the assessment of the
effectiveness of the self-organizing system functioning based on the
fractal-cluster approach has significant differences from the traditional
approach to this problem. Thus, we can conclude that the synthesis
of the FC theory and generalized thermodynamics of irreversible processes
allows us to determine the type of stability criterion for ``organisms''
placed far from the equilibrium state.

\section{Deterministic fractal-cluster theory}

{}\label{Det}

{}The aim of this study is to work out a resource management theory
of SOS based on the synthesis of fractal-cluster relations (FCR) and
non-equilibrium thermodynamics of I. Prigozhine.

{}The research of distribution based on the fractal-cluster approach
shows that the researched object is not decomposed, but is a ``black
box'', which corresponds to the principles and methodology of thermodynamics.
In the physical space the real object has the resources $X_{i}$,
necessary for its functioning, and the results of its activities.

\begin{figure}
\begin{centering}
{}\includegraphics[scale=0.8]{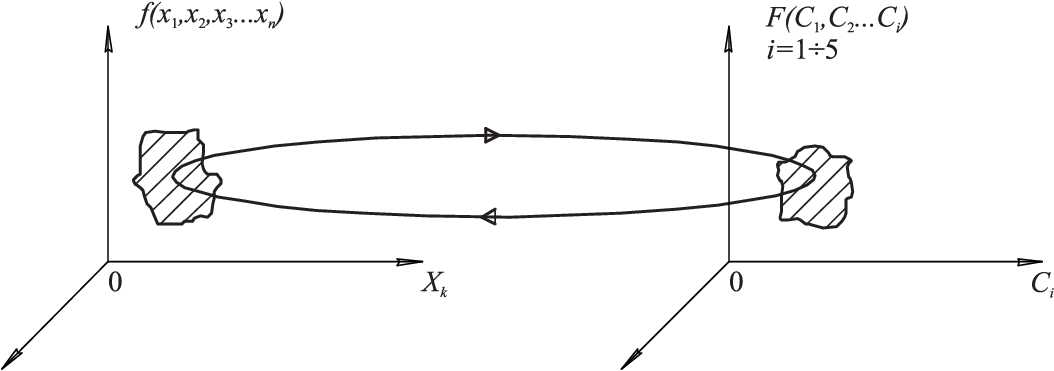}
\par\end{centering}
{}\raggedright{}\caption{Scheme of transfer from the physical space of the resources into the
fractal-cluster space. \label{figure5} }
\end{figure}

{}Upon transfer from the physical space of the external variable
in the five-dimensional fractal-cluster space, the decomposition and
classification of information about the object resources have been
accomplished on \fig{figure5}, i.e., the fractal-cluster structurization
of the information about the necessary resources for the SOS (energy,
transport, ecological, technological and informational supporting).
The universal thermodynamic apparatus in its informational interpretation
is convenient to use in this case.

{}

{}The laws and theorems of thermodynamics make it possible to analyze
the stability and the resource distributional efficiency of the SOS,
without additional empirical information.

{}The SOS fractal-cluster criteria are defined in a nontrivial way
by the construction of a fractal-cluster matrix (FCM). In the fractal-cluster
space, the cluster values $\left\{ C_{i}\right\} $ and subclusters
of any level are a positive value: $\overline{C}_{i}>0.$

{}The proposed theory is based on:

{}1. The axiom of the FCR universality (five-cluster structuring
of the SOS resource needs).

{}2. The assumption that clusters $\left\{ \overline{C}_{i}\right\} $
and subclusters of any level $\left\{ \overline{C}_{ij}\right\} ,\:\left\{ \overline{C}_{ijk}\right\} \ldots\left\{ \overline{C}_{i_{1}\ldots i_{n}}\right\} $
cannot take zero value: $\overline{C}_{i}>0,\:\overline{C}_{ij}\geq0,\:\ldots\overline{C}_{i_{1}\ldots i_{n}}\geq0$.

{}3. The assumption that the SOS effective area in the physical space
also corresponds to the effective functioning in the fractal-cluster
space. From the point of view of the FC theory, an effective area
of the SOS functioning is such a phase space of trajectories in which
low-intensity oscillations of the values of the FC criteria around
their reference values are realized. For self-organizing systems --
``biological organisms'' - this term (effective functioning) is
understood as a change in the distribution of resources of an organism
in the course of adaptation to environmental conditions. For self-organizing
systems -- ``social and economic systems'' -- this term means
a directed change in the distribution of system resources for the
stable dynamics of fractal-cluster criteria while simultaneously achieving
their extremeness.

{}4. The admission of the passive resource management model whith
delayed feedback.

{}The task of the resource distributional control can generally be
formulated as follows: $\left|u-u^{stab}\right|\rightarrow\min$,
where $u$ is a controlling function, $u^{stab}$ is a stable resource
distribution in the system, obtained on the basis of information and
thermodynamic methods (defined below).

{}The presented fractal-cluster theory includes:

{}1) V.P. Burdakov's FCR \cite{Burd};

{}2) the dynamical equations for the fractal-cluster system's evolution
\cite{GP};

{}3) the fractal-cluster criteria for the system's control efficiency
\cite{Burd,Volov1};

{}4) analysis of the complex self-organizing system's stability \cite{Burd,Volov1}.

\subsection[Fractal-cluster dynamic equations for the resource allocation]{Fractal-cluster dynamic equations for the resource allocation in
the self-organizind system}

{}From the point of view of the fractal-cluster approach, the description
of control resources in self-organizing systems is the redistribution
of resources corresponding to the achievement of the reference values
of clusters and subclusters fractal-cluster structure of SOS resources.

{}The fractal-cluster system controling can be discribed by equations:

{}
\begin{equation}
\overline{C}_{i_{1}\ldots i_{n}}\left(t\right)=F_{i_{1}\ldots i_{n}}\left(\overline{C}_{i_{1}\ldots i_{n}}^{0},\:f_{i_{1}\ldots i_{n}}\left(t\right)\right),\label{C_i_k}
\end{equation}
where $F_{i_{1}\ldots i_{n}}$ are given functions, associated with
non-simultaneity redistribution of resources in clusters, $f_{i_{1}\ldots i_{n}}\left(t\right)$
are control functions, $\overline{C}_{i_{1}\ldots i_{n}}^{0}=\overline{C}_{i_{1}\ldots i_{n}}\left(0\right)$.
For the function $F_{i_{1}\ldots i_{n}}$ which is, in a certain sense,
the possible values of a combination of clusters and subclusters,
three scenarios are possible: 1) management of all clusters and subclusters
occurs simultaneously (the beginning of resource transformation and
its end is the same for all clusters and subclusters); 2) the beginning
of resource transformation and its end is different for all clusters
and subclusters; 3) for some clusters and subclusters, there is a
simultaneous transformation of resources. For consistency, the function
$F_{i_{1}\ldots i_{n}}\left(\overline{C}_{i_{1}\ldots i_{n}}^{0},\:f_{i_{1}\ldots i_{n}}\left(t\right)\right)$
on the right-hand side of (\ref{C_i_k}) must satisfy the following
conditions.

{}1. Initial conditions

{}
\begin{equation}
F_{i_{1}\ldots i_{n}}\left(\overline{C}_{i_{1}\ldots i_{n}}^{0},\:f_{i_{1}\ldots i_{n}}\left(0\right)\right)=\overline{C}_{i_{1}\ldots i_{n}}^{0}.\label{F_i_k}
\end{equation}

{}2. For each (sub)cluster identified by the indices $\left(i_{1}\ldots i_{n}\right)$,
there is a moment of time $t_{i_{1}...i_{n}}^{final}$ such that

{}
\begin{equation}
\underset{t\rightarrow t_{i_{1}...i_{n}}^{final}}{\lim}F_{i_{1}\ldots i_{n}}\left(\overline{C}_{i_{1}\ldots i_{n}}^{0},\:f_{i_{1}\ldots i_{n}}\left(t\right)\right)\,=\overline{C}_{i_{1}\ldots i_{n}}^{ideal},\label{lim_F}
\end{equation}
where $\overline{C}_{i_{1}\ldots i_{n}}^{ideal}$ are ideal (sub)cluster
values.

{}3. Conserving the value of the total resource at any point in time
requires the fulfillment of the condition
\begin{equation}
\sum_{i_{1}=1}^{5}\ldots\sum_{i_{n}=1}^{5}F_{i_{1}\ldots i_{n}}\left(\overline{C}_{i_{1}\ldots i_{n}}^{0},\:f_{i_{1}\ldots i_{n}}\left(t\right)\right)=1.\label{sum_F}
\end{equation}

{}We consider the followiong simple structure of the equations (\ref{C_i_k})
which satisfy the conditions (\ref{lim_F})--(\ref{sum_F}) \cite{Volov1,Volov2}.

{}
\begin{equation}
\left\{ \begin{array}{l}
\overline{C}_{i}\left(\bar{t}\right)=\overline{C}_{i}^{0}+u_{i}\left(\varepsilon_{i},\,\bar{t}\right)\cdot\overline{C}_{i}^{0},\\
\overline{C}_{ij}\left(\bar{t}\right)=\overline{C}_{ij}^{0}+u_{ij}\left(\varepsilon_{ij},\bar{t}\right)\cdot\overline{C}_{ij}^{0},\\
\overline{C}_{ijm}\left(\bar{t}\right)=\overline{C}_{ijm}^{0}+u_{ijm}\left(\varepsilon_{ijm},\bar{t}\right)\cdot\overline{C}_{ijm}^{0},\\
\cdots\\
\overline{C}_{i_{1}\ldots i_{n}}\left(\bar{t}\right)=\overline{C}_{i_{1}\ldots i_{n}}^{0}+u_{i_{1}\ldots i_{n}}\left(\varepsilon_{i_{1}\ldots i_{n}},\bar{t}\right)\cdot\overline{C}_{i_{1}\ldots i_{n}}^{0}
\end{array}\right.\label{sys}
\end{equation}

\noindent {}at $0\leq\bar{t}\leq1$, where $\bar{t}=\dfrac{t}{t_{end}}$,
$t_{end}$ is end time of SOS management, where

{}
\begin{equation}
\left\{ \begin{array}{l}
u_{i}=\varepsilon_{i}f_{i}\left(t\right),\\
u_{ij}=\varepsilon_{ij}f_{ij}\left(t\right),\\
u_{ijm}=\varepsilon_{ijm}f_{ijm}\left(t\right),\\
\cdots\\
u_{i_{1}\ldots i_{n}}=\varepsilon_{i_{1}\ldots i_{n}}f_{i_{1}\ldots i_{n}}\left(t\right),
\end{array}\right.\label{sys_1}
\end{equation}
where
\[
\left\{ \begin{array}{l}
\varepsilon_{i}=\left(\dfrac{C_{i}^{0}}{C_{i}^{id}}-1\right),\\
\varepsilon_{ij}=\left(\dfrac{C_{ij}^{0}}{C_{ij}^{id}}-1\right),\\
\varepsilon_{ijm}=\left(\dfrac{C_{ijm}^{0}}{C_{ijm}^{id}}-1\right),\\
\ldots\\
\varepsilon_{i_{1}\ldots i_{n}}=\left(\dfrac{C_{i_{1}\ldots i_{n}}^{0}}{C_{i_{1}\ldots i_{n}}^{id}}-1\right)
\end{array}\right.
\]
and

{}
\[
\begin{array}{c}
f_{i}\left(0\right)=f_{ij}\left(0\right)=\ldots=f_{i_{1}\ldots i_{n}}\left(0\right)=0,\\
f_{i}\left(1\right)=f_{ij}\left(1\right)=\ldots=f_{i_{1}\ldots i_{n}}\left(1\right)=1.
\end{array}
\]

\noindent {}Here $u_{i},\:u_{ij},\:u_{ijm},\:\ldots,\:u_{i_{1}\ldots i_{n}}$
are controlling functions for clusters and subclusters of the first,
second and $(n-1)$ levels, $\overline{C}_{ij}^{ideal},\:\overline{C}_{ijm}^{ideal},\:\ldots,\:\overline{C}_{i_{1}\ldots i_{n}}^{ideal}$
are ideal relative subcluster's values of the first, second, ...,
$(n-1)$ levels, and $\overline{C}_{ij}^{0},\:\overline{C}_{ijm}^{0},\:\ldots,\overline{C}_{i_{1}\ldots i_{n}}^{0}$
are initial relative respective subclusters values, we assume that
$f_{i}\left(t\right)=f_{ij}\left(t\right)=\ldots=f_{i_{1}\ldots i_{n}}\left(t\right)=f\left(t\right)$,
where $f\left(t\right)$ is a monotonous differentiable function $0\leq f\left(t\right)\leq1$,
whose form is either given or found from the additional stability
conditions. Equation (7) is an analogue of the conservation law for
the fractal-cluster system.

\subsection[Entropic-cluster method of analysis and management]{Entropic-cluster method of analysis and management for self-organizing
systems}

{}The control optimizational methods which have been proposed in
\cite{Stakhov0,Stakhov1,Stakhov2,GGZ} are based on intuitive or rigidly
formalized concepts and analogies. In connection with the above, it
is logical to formulate a criterion of control efficiency for the
FCM on the basis of fundamental principles of stable states thermodynamics.

{}Let us consider the FCM of ideal states (two-dimensional case $N=2$,
\tabl{tab2}).

{}The first row and the first column resources of an ideal matrix
give quantitative information about the overall share of system energy
resources, which is $0.615$, this being the major determinant of
the system's functioning effectiveness.

{}The relationship between elements of the FCM and the information
entropy $H$ allows us to find a criterion for FCM controlling for
the purpose of optimal evolution from the nonideal state of the system
(non-ideal FCM) into the perfect condition -- (ideal FCM), so that,
the sum of FCM elements of the first column and the first row (\tabl{tab2})
acordind (\ref{H_0}) goes into their ideal value, that is, the ``Golden
section'' entropy value. For the $n$-levels of FCM the fractal-cluster
entropy has the form (\ref{H_n}). Thus, we introduce the mathematical
measure -- an entropy (or quasientropy), based on a generalization
of experimental data on the evaluating systems \cite{Burd} and the
FCM structure.

{}The proposed expression of the fractal-cluster entropy is, in terms
of value the share of the system total resources used to satisfy its
all energy needs. The structure of the FCM of complex system is a
fractal. Any elements of the one are chains of repeating subclusters,
self-similar in their structure. As it is known, a fractal image is
obtained by the iterative processes \cite{Mand}. The elementary iterative
process is Fibonacci numbers.

{}It turned out that the key to the FCM controlling is the famous
Fibonacci numbers range $\left\{ U_{n}\right\} =\left\{ 0,\:1,\:1,\:2,\:3,\:5,\:8,\:13,\ldots\right\} $,
in which each successive number is the sum of the previous two. A
remarkable property of Fibonacci numbers is that with numbers increasing
the ratio of two adjacent numbers is asymptotically close to the exact
proportion of the ``Golden section'' \cite{Stakhov0,Stakhov3,S_R}:

{}
\begin{equation}
\lim_{n\rightarrow\infty}\frac{U_{n}}{U_{n+1}}=H_{0}\approx0.618.\label{U_U}
\end{equation}

\begin{figure}[H]
\begin{centering}
{}\includegraphics[scale=0.8]{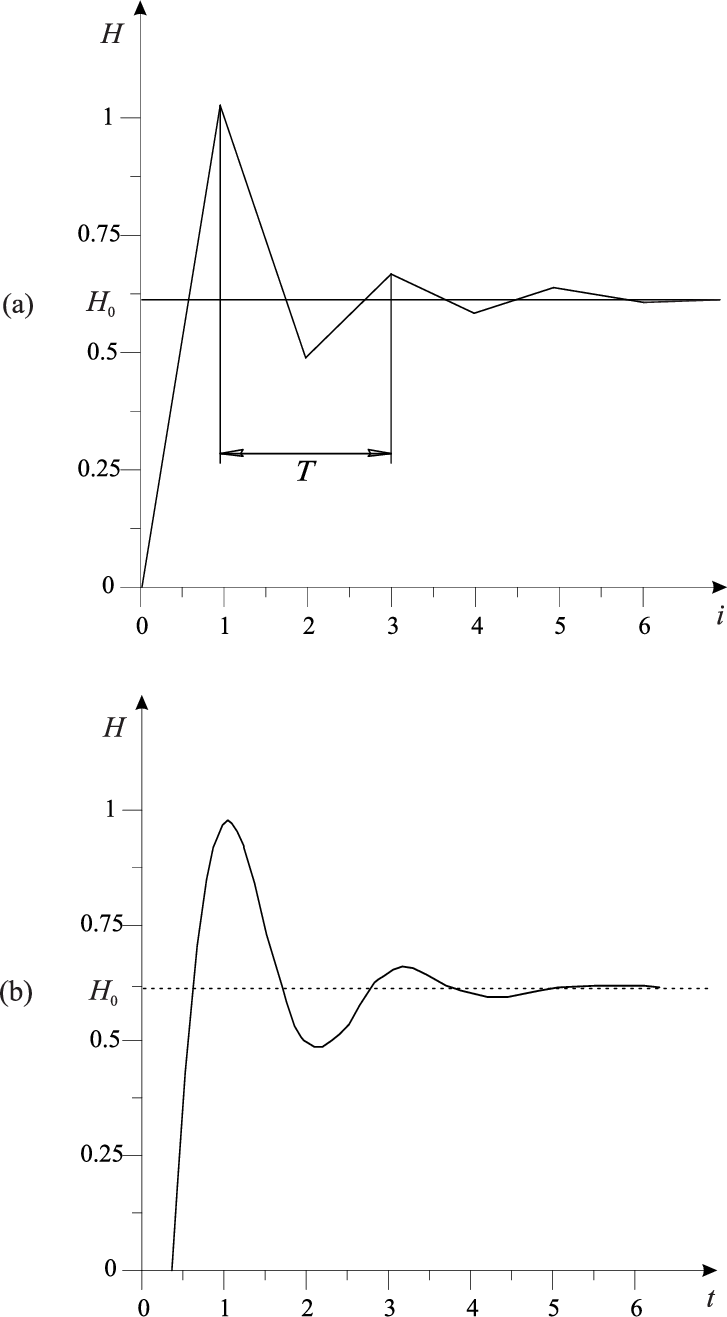}
\par\end{centering}
{}\centering{}\caption{Iteration of Fibonacci numbers: $i\:$ is an iteration number, $T=2$
is a period, $H_{0}\approx0.618\:$ is the \textquotedblleft Golden
section\textquotedblright{} (a); the approximation of Fibonacci number's
iterations (b). \label{figure6} }
\end{figure}

{}In connection with the above, there is a hypothesis about the optimal
management of the FCM by means of the Fibonacci numbers. To control
the FCM the approximation of the Fibonacci numbers iterations \cite{S_R}
is used. Here iteration corresponds to time intervals that are multiple
to the period of entropy oscilations, i.e. the approximation of the
Fibonacci numbers iterations is a template for the matrix management
(\fig{figure6}) $\left\{ u_{ij}\right\} $.

{}

{}The beginning control of the cluster evolution and the end of the
ones are realized at the same time $\overline{t}_{ij}^{0}=\overline{t}_{ji}^{0}=const$
and $\overline{t}_{ij}^{fin}=\overline{t}_{ji}^{fin}=const$. It takes
the form:

{}
\begin{equation}
u_{ij}=\left(\dfrac{\overline{C}_{ij}^{0}}{\overline{C}_{ij}^{id}}-1\right)\cdot f\left(\overline{t}-\overline{t}_{0}\right).\label{u_ij}
\end{equation}
Function $f\left(\overline{t}-\overline{t}_{0}\right)$ satisfies
the conditions (\ref{sys_1}).

{}An approximation of Fibonacci numbers iterations (\fig{figure6}b)
gives the following expression:

{}
\begin{equation}
f\left(\overline{t}-\overline{t}_{0}\right)=\frac{H}{H_{0}}=1+H_{0}\cdot\exp\left(-\alpha\left(\overline{t}-\overline{t}_{0}\right)\right)\times\cos\left(\pi\left(\overline{t}-\overline{t}_{0}\right)+\varphi_{0}\right),\:\overline{t}_{ij}^{fin}=\overline{t}_{ji}^{fin}=const_{2}\label{f_t}
\end{equation}
at the initial stage $\varphi_{0}=0,\,H_{0}=0.618,\,\alpha=1,05,\,\overline{t}_{0}=1.$
The expression (\ref{f_t}) fails to satisfy the initial conditions
at $\overline{t}=t_{0}$. To satisfy the second boundary condition
we introduce a new control function $u^{\star}$ at the time interval
from zero to some $t$:

{}
\begin{equation}
u^{\star}\left(\overline{t}-\overline{t}_{0}\right)=1-\exp\left(-\beta\left(\overline{t}-\overline{t}_{0}\right)\right),\label{u_star}
\end{equation}
and make sewing of solutions for $U\left(\overline{t}-\overline{t}_{0}\right)$
we receive:

{}
\begin{equation}
\left\{ \begin{array}{c}
u_{1}^{\star}\left(\overline{t}-\overline{t}_{0}\right)=f\left(\overline{t}-\overline{t}_{0}\right),\\
\left(u^{\star}\left(\overline{t}-\overline{t}_{0}\right)\right)^{\prime}=f^{\prime}\left(\overline{t}-\overline{t}_{0}\right),
\end{array}\;\textrm{at}\right.\overline{t}=\overline{t}_{sewing}.\label{u_sys}
\end{equation}
It is evident that the control $u^{\star}\left(\overline{t}-\overline{t}_{0}\right)$
satisfies condition (\ref{sys_1}) at $\overline{t}=\overline{t}_{0}$.
After simple transformations we get a system of the transcendental
equations:

{}
\begin{equation}
\left\{ \begin{array}{c}
\beta=\alpha-\dfrac{\log\left(-H_{0}\cos\left(\pi\left(\overline{t}_{sewing}-\overline{t}_{0}\right)\right)\right)}{\overline{t}_{sewing}-\overline{t}_{0}}\\
\ln\left(-\dfrac{1}{H_{0}\cos\left(\pi\left(\overline{t}_{sewing}-\overline{t}_{0}\right)\right)}\right)=tg\left(\pi\left(\overline{t}_{sewing}-\overline{t}_{0}\right)\right)
\end{array}\right.,\label{beta_sys}
\end{equation}
where $\cos\left(\pi\left(\overline{t}_{sewing}-\overline{t}_{0}\right)\right)_{0}<0$.
Numerically the index $\beta$ is determined from the equations (\ref{beta_sys})
the value of $t_{sewing}$ and it turned out that $\overline{t}-\overline{t}_{0}\approx1.19;\;\beta\cong1.53$.

{}
\begin{figure}[H]
\begin{centering}
{}\includegraphics[scale=0.5]{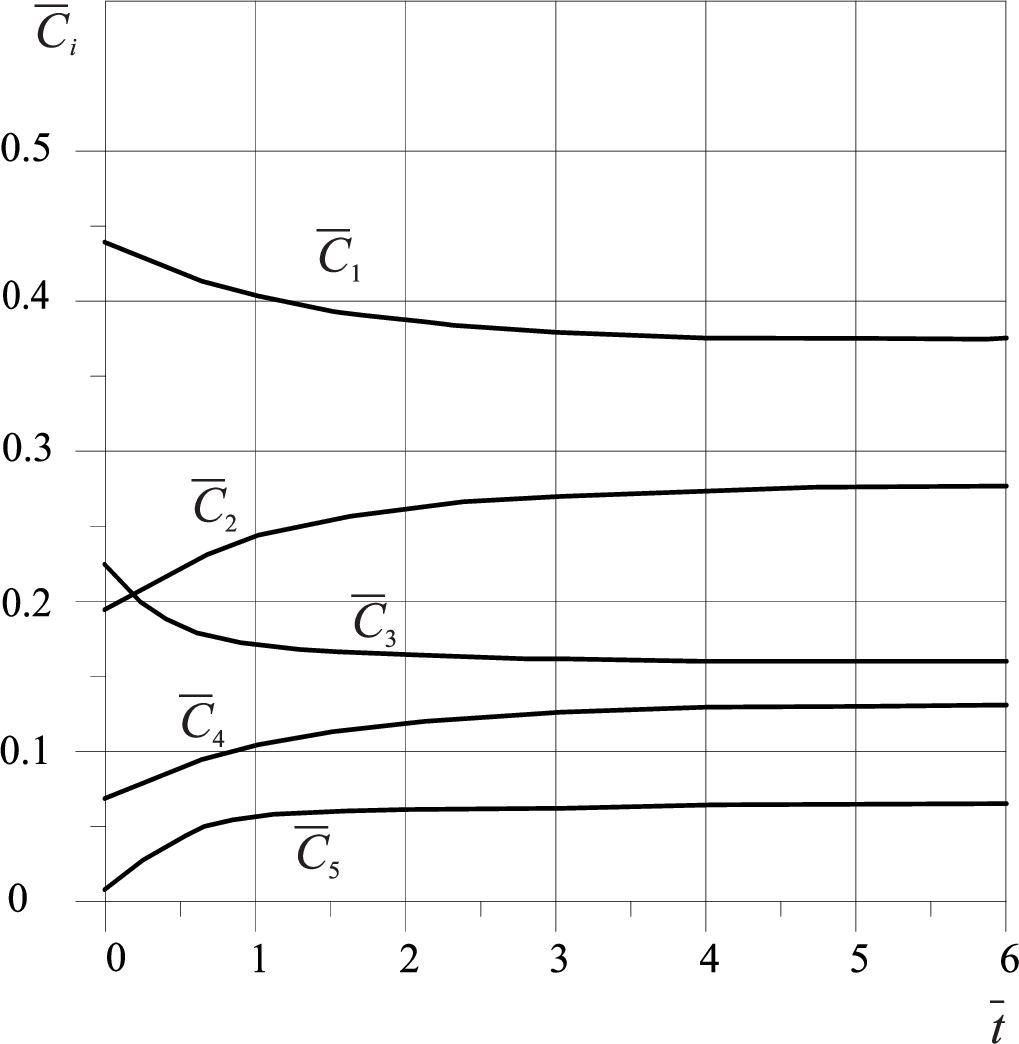}
\par\end{centering}
{}\centering{}\caption{Evolution of $SOS$ clusters according to Fibonacci numbers and condition
of the sewing (\ref{u_star}). \label{figure7} }
\end{figure}

{}\fig{figure7} shows the evolution of clusters controlled by (\ref{f_t},\ref{u_star}).
As seen from \fig{figure6}, these are stable (convex trajectory
$H,\;\dfrac{d^{2}H}{dt^{2}}<0$) and unstable (concave trajectory
$H,\;\dfrac{d^{2}H}{dt^{2}}>0$), corresponding to the obtained entropy-cluster
solution which is based on the Fibonacci numbers. In this connection
the following hypothesis was suggested: if the structural entropy
of low intensity then the regimes where $\delta H\ll H_{0},\;\dfrac{dP}{dt}>0$
is a mode of functional instability, which is an attribute of any
developing self-organizing system. In the unstable regimes $\left(\dfrac{d^{2}H}{dt^{2}}>0\;\textrm{and}\:\delta H\sim H_{0}\right)$
there exists an anomalous structural instability, which means these
is a serious crisis of structural processes in the SOS. These modes
of SOS operation are discussed in detail in (\tabl{tabnew}).

{}The criteria $H$, \emph{{}$D$}{} and $\dfrac{d^{2}H}{dt^{2}}$
make it possible to determine the necessary and sufficient conditions
of optimal resource distribution in $SOS$ in a static state (\tabl{tab3}).

{}
\begin{table}[H]
{}\caption{ Necessary and sufficient conditions of SOS optimal state.}
\label{tab3} \centering{}%
\begin{tabular}{|>{\raggedright}m{4cm}|>{\raggedright}m{4cm}|>{\raggedright}m{4cm}|}
\hline
{}A necessary condition for the optimal functioning of the economic
system  & {}A sufficient condition for the optimal functioning of the economic
system  & {}Notes\tabularnewline
\hline
{}$H\rightarrow H_{0},$ or $\left|H-H_{0}\right|\rightarrow0$  & {} $\dfrac{d^{2}H}{dt^{2}}<0$, $D\rightarrow D^{\max}$  & {}There is complete information on SOS (accurate estimate)\tabularnewline
\hline
{}$H\rightarrow H_{0},$ or $\left|H-H_{0}\right|\rightarrow0$  & {}$\eta^{\Sigma}\rightarrow1$  & {}There is no complete information on SOS (rough estimate)\tabularnewline
\hline
\end{tabular}
\end{table}

{}Let us consider the problem of stability of transition trajectory
of a complex system from arbitrary in ideal condition in accordance
with the fractal-cluster theory basic provisions. Obviously, both
stable and unstable trajectories of system transformation can go through
the one point in the phase plane of the entropy - time $\left(H-t\right)$
in terms of the fractal-cluster theory (\fig{figure7}).

{}Let us also consider a fractal-cluster structure of a complex system
is placed in a state close to thermodynamic equilibrium, that is,
while analyzing evolution it is possible to use the linear thermodynamics
of non-equilibrium processes. In accordance with this fact, we can
use the theorem of minimum entropy production \cite{GP}. For simplicity,
we consider a symmetric FCM of a topological structure, then the SOS
entropy is determined by (\ref{H_2}).

{}Using I. Prigogine's theorem of minimum entropy production \cite{GP},
we define the function $f\left(\overline{t}\right)$ from the condition
of neutral stability:

{}
\begin{equation}
\frac{dP}{d\overline{t}}=0,\;\textrm{where}\;P=\frac{dH}{d\overline{t}}.\label{dP}
\end{equation}

{}The expression of transformation function $f\left(\overline{t}\right)$,
which executes the transition trajectory of a fractal-cluster system
from the arbitrary into an ideal state, in the trajectory of neutral
stability is the following:

{}
\begin{equation}
f\left(\overline{t}\right)=\left\{ \exp\left(\alpha\right)-1\right\} ^{-1}\left(\exp\left[\alpha\cdot\overline{t}\right]-1\right).\label{f_t_exp}
\end{equation}

{}In the general case of non-zero right-hand side in the expression
for the entropy production we obtain the following expression for
the transformation function $f\left(\overline{t}\right)$:

{}
\begin{equation}
f\left(\overline{t},\varepsilon\right)=-\frac{\varepsilon}{\alpha}\overline{t}+\left(1+\frac{\varepsilon}{\alpha}\right)\cdot\left(\exp\left(\alpha\right)-1\right)^{-1}\left(\exp\left(\alpha\cdot\overline{t}\right)-1\right).\label{f_t_eps}
\end{equation}
The expression for the function $f\left(\overline{t},\varepsilon\right)$
conforms to the following qualitatively different transformation modes
of the topological fractal-cluster structure of a complex system from
the non-ideal to an ideal state:

{}
\begin{equation}
\left\{ \begin{array}{l}
\varepsilon=0\textrm{ - transformation of a complex system by the trajectory of a neutral stability,}\\
\varepsilon>0\textrm{ -\textrm{ unsustainable trajectory of a complex system transformation,}}\\
\varepsilon<0\textrm{\textrm{ - steady trajectory of a complex system transformation}}.
\end{array}\right..\label{eps}
\end{equation}

\begin{figure}[H]
{}\centering{}\includegraphics[scale=0.6]{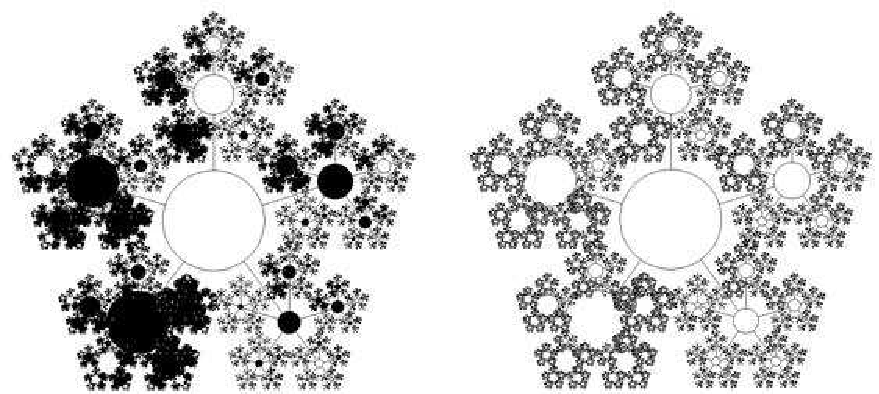}\caption{Evolution the FC resource structure for the case of a six-level FCM:
$\varphi=72^{\circ}$, $\:t=0\:$ (left), $\varphi=72^{\circ}$, $\:t=1\:$
(right). \label{figure8} }
\end{figure}

\begin{figure}[H]
{}\centering{}\includegraphics[scale=0.4]{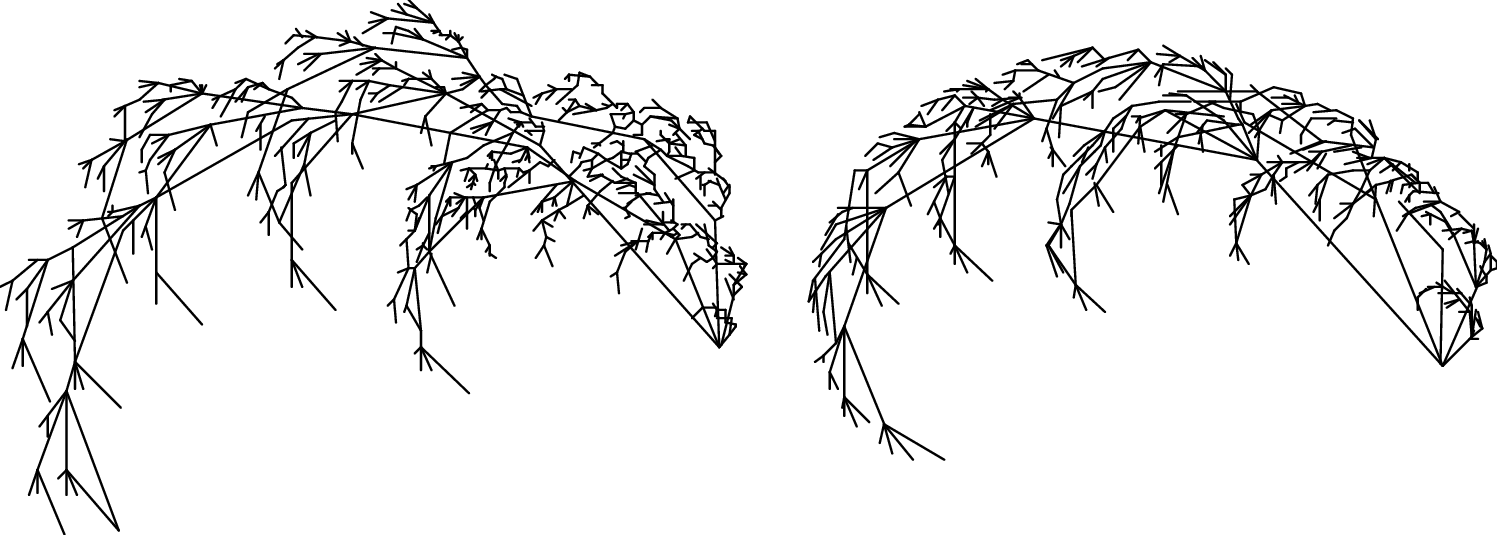}\caption{Evolution the FC resource structure for the case of a six-level FCM:
$\varphi=20^{\circ},\:t=0\:$ (left), $\varphi=20^{\circ},\:t=1\:$
(right). \label{figure9} }
\end{figure}

{}\fig{figure8} and \fig{figure9} show the various scenarios
of the FCM topological structures for different time stages of FCM
evolution to its ideal value. Here a graphic illustrations of topological
fractal-cluster structures for various scenario and schemes of clusters
and subclusters distributions are presented. It is shown an evolution
from an initial (nonideal) state to an final (ideal) state of the
FC structure (for $n=6$ and for rotation angles $\varphi=20^{\circ}$
(\fig{figure8}) and $\varphi=72^{\circ}$(\fig{figure9})) for
tree image stuctures. Circle squares (\fig{figure8}) and tree branches
lengthes (\fig{figure9}) are equal corresponding cluster (subcluster)
values.

{}

{}

\section{Foundations of probabilistic fractal-cluster theory }

{}\label{Prob}

{}The second part of this article has presented the determined fractal-clusters
theory foundations. This one allows us to receive the criterion's
estimations about the functioning of the economic system. But the
real self-organizing systems have a stochastic behavior. The parameters
have a fluctuational character. That's the reason why real $SOS$
characteristics are more preferable for research by the probabilistic
equations.

{}Let's determine the phase space of the one level FC system as the
set $Q\equiv\left\{ Q_{a}\right\} $, $Q_{a}=\left(\overline{C}_{1}^{(a)},\:\overline{C}_{2}^{(a)},\:\overline{C}_{3}^{(a)},\:\overline{C}_{4}^{(a)},\:\overline{C}_{5}^{(a)}\right)$
under $\overset{5}{\underset{i=1}{\sum}}C_{i}^{(a)}=1$ and the index
$a$ identifies the elements of the set $Q$. We will be to learn
the discrete approximation $Q$. For the discretization of $Q$ we
have to divide all possible cluster values on finite elements so on
the set $Q$ will be to have finite number of elements $N$. Under
$F\left(Q_{a},t\right)$ will be to understand probability of the
system placing in the state $Q_{a}$ at the time $t$.

{}We will be to discribe of evolution of the distribution function
by ordinary differential equations system like the Kolmogorov--Feller's
one \cite{Gard}:

{}
\begin{equation}
\frac{dF\left(Q_{a},t\right)}{dt}=\sum_{a'}\left[W\left(Q_{a},Q_{a'}\right)F\left(Q_{a'},t\right)-W\left(Q_{a'},Q_{a}\right)F\left(Q_{a},t\right)\right],\label{dF}
\end{equation}

{}We can research the approximation in which all states with equal
$D$ have the equal probability. In this case the system will be described
by the function of distribution

{}
\[
f\left(D,t\right)=N\left(D\right)F\left(Q_{a},t\right),
\]
where $N\left(D\right)$ is a number of states with fixed value of
$D$. In this case the equation (\ref{dF}) has the following form

{}
\begin{equation}
\frac{df\left(D,t\right)}{dt}=\sum_{D'}\left[w\left(D,D^{\prime}\right)f\left(D^{\prime},t\right)-w\left(D^{\prime},D\right)f\left(D,t\right)\right],\label{ME}
\end{equation}
where $w\left(D,D^{\prime}\right)=N\left(D\right)W\left(Q,Q^{\prime}\right)$
is a probability of transition per unit of time from the group of
states $Q'$ with fixed value of $D^{\prime}$ to the group of states
$Q$ with fixed value of $D$.

{}For the first, the most interest presents the stationary solution
of these equations (\ref{ME}). The stationary solution of these equations
(\ref{dF}) under given values of the $w\left(D,D^{\prime}\right)$
presents itself the standard task. However, obtainig of the stationary
solution to these equations is possible by using numerical calculation.
From the definition of the $D$ criterion (\ref{D_crit}) is determined
larger the limit values of the clusters $\left\{ C_{i}\right\} :$

{}
\begin{equation}
a_{i}\leq c_{i}\leq b_{i}\:i=1\div5\label{c_restr}
\end{equation}

{}
\[
\left\{ a_{i}\right\} =\left\{ \begin{array}{c}
0.01\\
0.01\\
0.01\\
0.01\\
0.01
\end{array}\right\} ,\:\left\{ b_{i}\right\} =\left\{ \begin{array}{c}
0.95\\
0.89\\
0.70\\
0.60\\
0.31
\end{array}\right\}
\]

{}For every interval of the allowed cluster values we can divide
these intervds into $n$ shares for each cluster. The possibility
of the functioning of the complex fractal-cluster system ``an organism''
is realized under the execution resource conservation law (\ref{conserv}).

\begin{figure}[H]
\begin{centering}
{}\centering{}\includegraphics{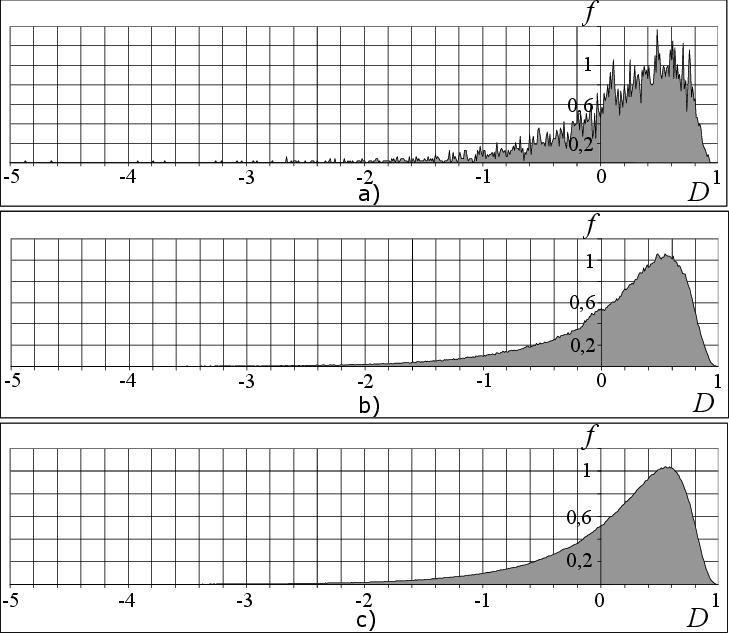}
\par\end{centering}
\caption{Distribution of the SOS probability density $f(D)$: $n=10$ (a),
$n=30$ (b), $n=50$ (c). \label{figure10} }
\end{figure}

\begin{figure}[H]
\centering{}{}\centering{}\includegraphics[scale=0.8]{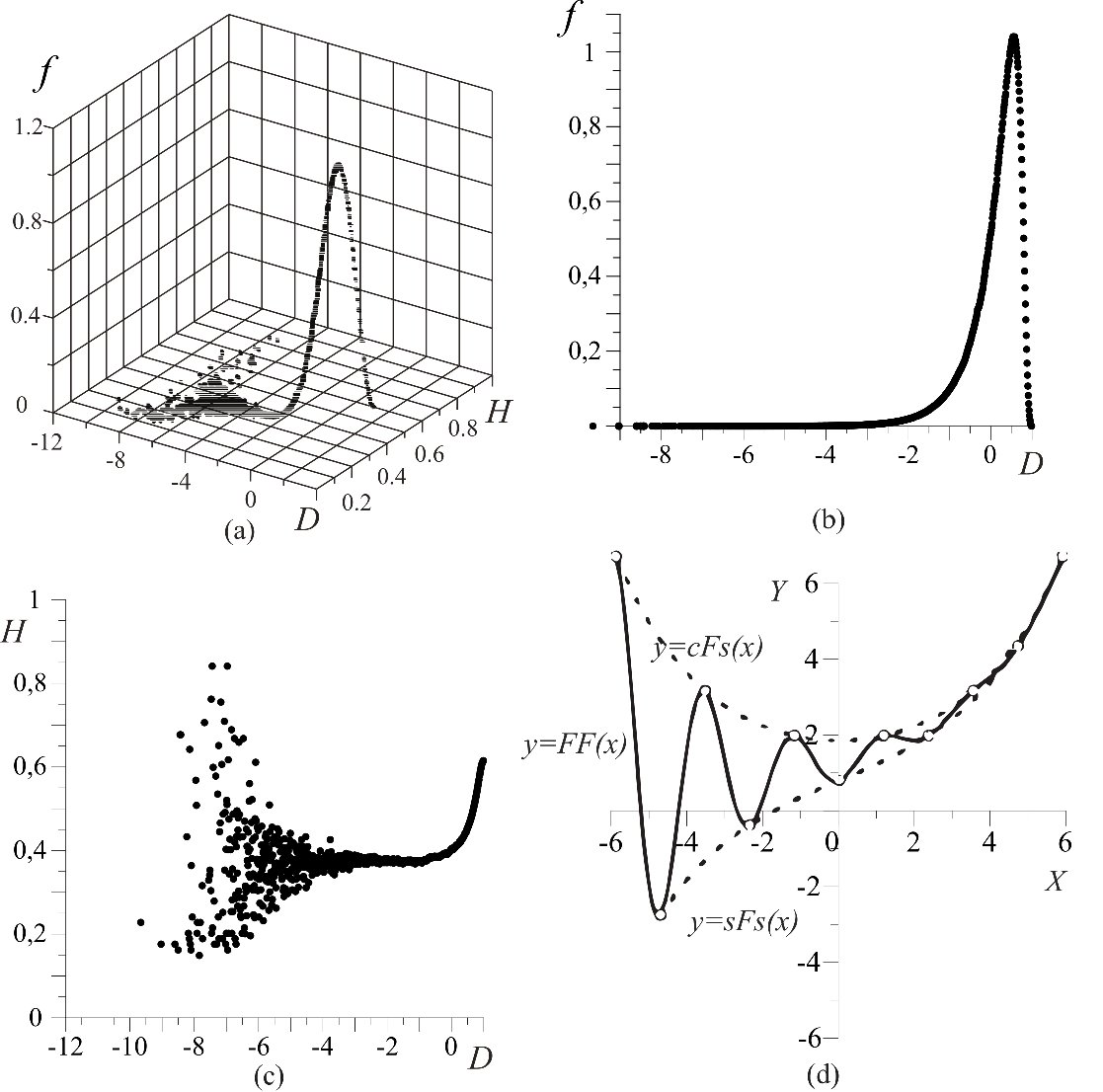}\caption{Probabilistic characteristics of FC system states: the dependence
$f\left(H,D\right)$, where $D$ is the efficiency criterion of recourse
distribution, $f$ is a probability density, $H$ is a fractal-cluster
entropy (a); the dependence $f\left(D\right)$(b); the dependence
$H\left(D\right)$ (c); the \textquotedblleft Gold Shofar\textquotedblright{}
\cite{St_GS} (d). \label{figure11} }
\end{figure}

{}That is why only some combinations of the valid values clusters
$\left\{ C_{i}\right\} $can be realized. In \fig{figure10} the
distributions of the possible states under the various divisions of
the valid cluster intervals have been shown. Beginning with the $n>50$
($n$ is a numeral of the cluster's dividing interval) the picture
of these distributions refrain constant as shown on \fig{figure11};
we can also see the fractal-cluster distribution of the probability
density as a function of the \emph{{}$D$}{}, $H$ criteria.

{}

{}

{}The curve cone of the fractal-cluster states which looks like the
``Golden Shofar'' (the projection of the surface Shofar on the plane
$XOY$ \cite{Stakhov0}) can be seen in the plate space $\left(H-D\right)$
in \fig{figure11}.

{}This similarity between the fractal-cluster theory and the topology
of the mathematic theory of the ``Golden Section'' \cite{Stakhov0,Stakhov1,Stakhov2,GGZ,Stakhov3,S_R}
has a deep correlation.

{}The obtained distribution (\fig{figure11}b) can not be described
by the well-known probabilistic distributions. In the phase fractal-cluster
space each value of the \emph{{}$D$}{}-criterion (\fig{figure11}b)
corresponds to a big number of the cluster's combinations which satisfy
the resource conservation law (\ref{conserv}), (\ref{sum_F}).

\begin{figure}[H]
\centering{}{}\centering{}\includegraphics[scale=0.4]{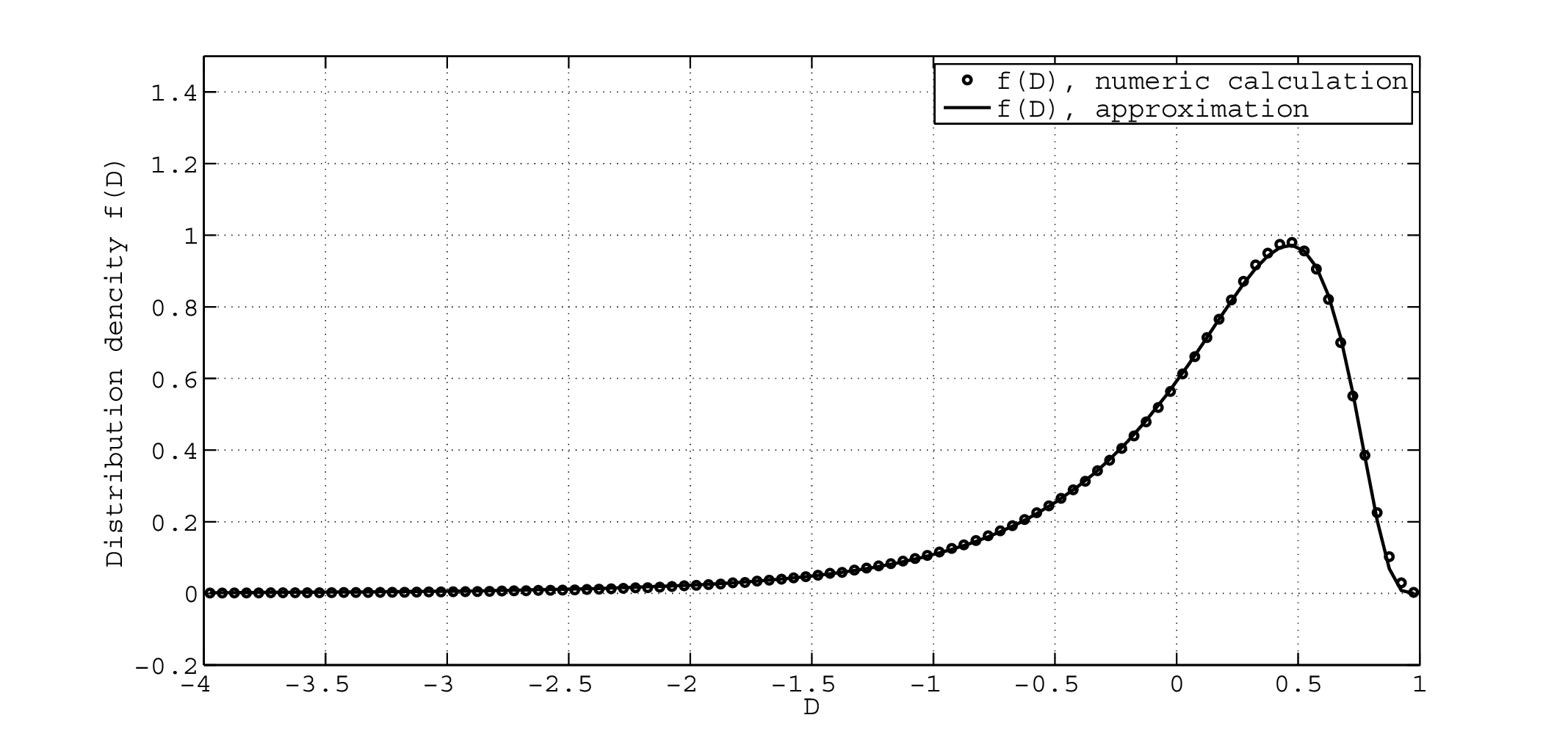}\caption{Lognormal approximation of the fractal-cluster probability dencity
$f\left(D\right)$. \label{figure11-1} }
\end{figure}

{}An approximation of the dependence $f-D$ (\fig{figure11}b) has
the following lognormal form:

{}
\begin{equation}
f\left(D\right)=\dfrac{1}{\sqrt{2\pi}\sigma\left(1-D\right)}\exp\left(-\dfrac{\left(\log\left(1-D\right)-\log a\right)^{2}}{2\sigma^{2}}\right),\label{P_D}
\end{equation}
where $a\approx0.77$, $\sigma\approx0.63\sim H_{0}$ (see \fig{figure11-1})

{}

{}Each cluster's combination is equally possible under the fixed
value of the $D$-criterion (\fig{figure11}b). For a reduction of
the cluster's combinations number we can use the thermodynamic method
of the minimum free energy principle for each value of the \emph{{}$D$}{}-criterion.
This gives us a decrease of intervals of the valid cluster's values
and this is very important for the real self-organizing system functioning
prediction. Besides the thermodynamic method for decreasing of the
cluster's interval we can use the probabilistic method.

{}We are interested in a stationary solution of the one. That's why
we can take the method of ``the local balance'' for each fractal-cluster
combination.

\begin{figure}[H]
\centering{}\centering{}{}\centering{}\includegraphics[scale=0.8]{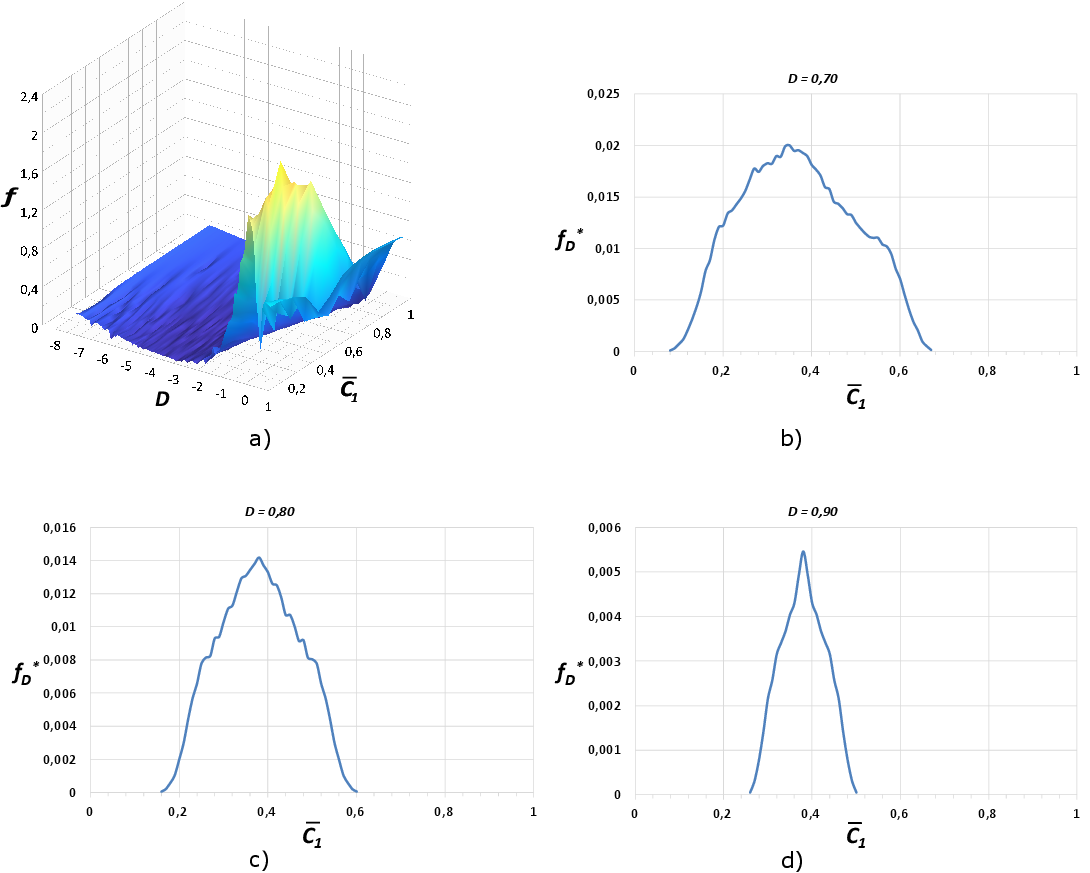}\caption{Probability states of the SOS in the fractal-cluster space, a) --
the probability density space $f$ of the FC system\textquoteright s
states, b), c), d) -- the reduced probability density $f_{D}^{\ast}$
as a function of $\overline{C}_{1}$ under fixed values of $D$. \label{figure11-2}}
\end{figure}

\begin{figure}[H]
\centering{}\centering{}{}\centering{}\includegraphics[scale=0.8]{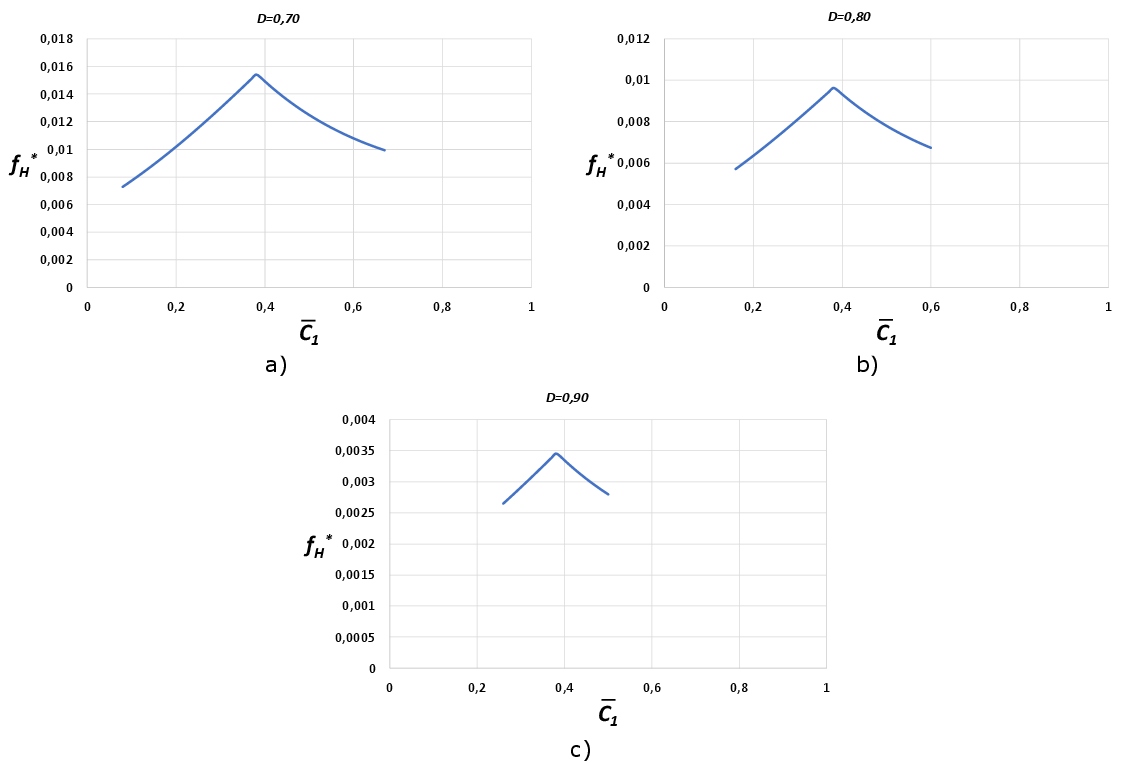}\caption{a), b), c) -- the reduced probability density $f_{H}^{\ast}$ as
a function of $\overline{C}_{1}$ under fixed values of $D$. \label{figure11-3}}
\end{figure}

Direct calculations of the probability density of the number of combinations
$\overline{C}_{i}$ as a function $f\left(\overline{C}_{1},D\right)$
and the reduced probability density $f_{H}^{\ast}$ of the values
of the energy cluster $\overline{C}_{1}$ at a fixed value of criterion
$D$ are shown in \fig{figure11-2} a, b, c, d. It can be seen from
\fig{figure11-2} b, c, d. that with an increase in the efficiency
of resource allocation in the SOS (an increase in criterion $D$ from
$0.7$ to $0.9$), the highest probability density of the number of
cluster combinations is achieved at the reference value of the energy
cluster $\overline{C}_{1}^{ideal}=0.38$ (38\%) of the total resource.
The following relation is used to estimate the reduced probability
density $f_{H}^{\ast}$ of entropy fluctuations:

{}
\begin{equation}
f_{H}^{\ast}=A\exp\left(-\dfrac{\left|H-H_{0}\right|}{H_{0}}\right),\label{f_sim}
\end{equation}
where $H$ is a fractal-cluster entropy, $H_{0}$ is the entropy of
the ``Golden Ratio'', $A$ is a normalization constant. This reduced
probability density ratio makes it possible to determine the range
of oscillations of the entropy trend $H$ relative to the entropy
of the ``Golden ratio'' $H_{0}$ (\ref{f_sim}), considering the
limitations for entropy trend oscillations (\tabl{tabnew} and \fig{figure11-2}
-- \fig{figure11-3}).

Calculations of the number of combinations of $\overline{C}_{i}$
cluster values ($i=1,\ldots,5$) in the fractal-cluster space show
that for fixed values of criterion $D>0.7$, the mathematical expectations
of $\overline{C}_{i}$ cluster values exactly correspond to their
reference values (\fig{figure1}). The standard deviation of cluster
values decreases with increasing criterion $D$. For example, with
the same interval of change of $C_{1}$ in the range $\left(0.35\div0.41\right)$
and $D=0.8$, we get $\overline{C}_{1}=38.0\pm2$, $\overline{C}_{2}=27.0\pm7.08$,
$\overline{C}_{3}=15.9\pm6.35$, $\overline{C}_{4}=12.98\pm5.44$,
$\overline{C}_{5}=6.18\pm2.98$ (in percent) and at $D=0.9$ the cluster
values and their standard deviations take the following values $\overline{C}_{1}=38.0\pm1.86$,
$\overline{C}_{2}=27.0\pm4.44$, $\overline{C}_{3}=16.0\pm3.95$,
$\overline{C}_{4}=13.0\pm2.72$, $\overline{C}_{5}=6.0\pm1.31$. Thus,
the values of mathematical expectations $M$ of combinations of clusters
$C_{i}$ obtained as a result of calculations, as well as their most
probable values in the fractal-cluster space, represent a probabilistic
confirmation of the reference (ideal) values of clusters obtained
experimentally \cite{Burd}.

\section[Applications of the fractal-cluster theory]{Applications of the fractal-cluster theory for the resource distribution
analysis and controlling in the socio-economical and biological systems }

{}\label{App}

\subsection{Economic systems}

\begin{figure}[H]
{}\centering{}\includegraphics[scale=0.7]{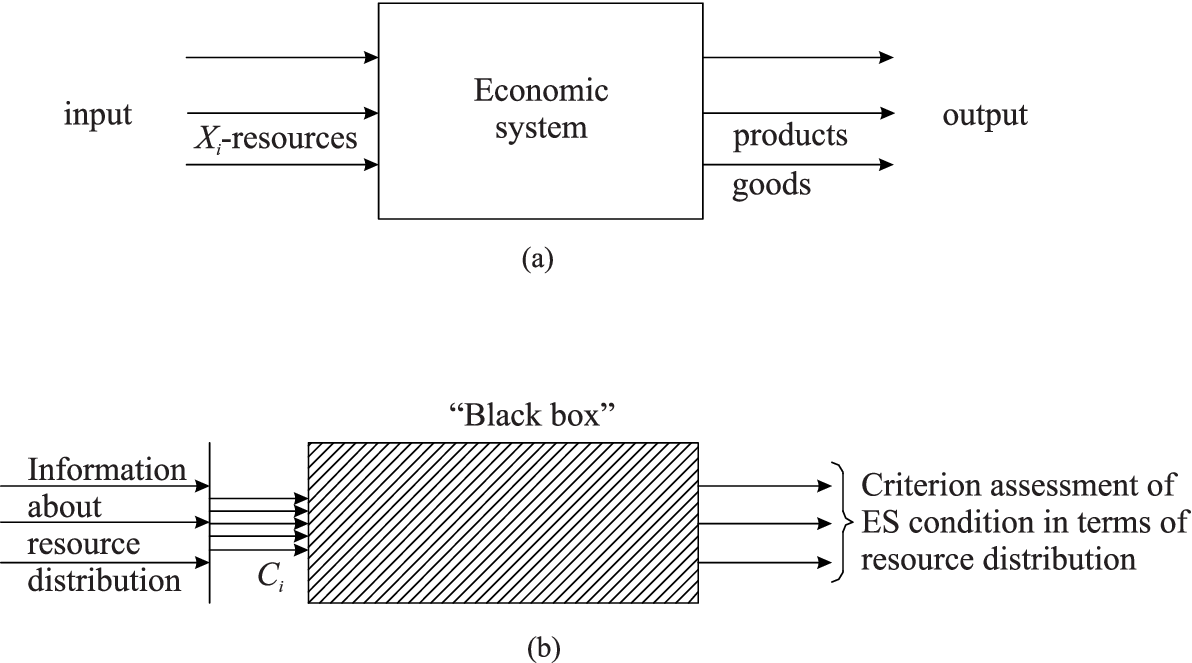}\caption{Diagram representing the economic system in the traditional (a) and
fractal-cluster (b) interpretation. \label{figure13} }
\end{figure}

{}In \fig{figure13} we have presented the principle defferences
between the traditional economic and fractal-clusters models.

{}In the traditional economic system (ES), ``input-output'' analysis
as a rule has known the price of products of the one. In this case,
for example, the Leontiev's model is used \cite{Leo}. In traditional
economic ``input-output'' system's model under known inputs resources,
$x_{i}$ is the determined output product price (a prolife etc.).
Using the fractal-cluster models we receive an estimation of the resource
utilization and distribution efficiency on the fractal-cluster criteria
basis in ES.

{}

{}It is well-known that the problem with inner resource distribution
in economic systems have quickly lead to crisis situations in the
financial, productional and other activites of these systems.

{}Therefore, by using the fractal-cluster theory, we can obtain the
special additional information about the resource distribution economic
system's functioning in advance. It allows us to introduce the correction
in the resource distribution economical system.

{}The result of informational-thermodynamic analysis of resourse
distribution in economical systems based on the fractal-cluster models
makes it possible to formulate a new generalized criterion to optimize
the economical system management is a optimal control of an economical
system in terms of fractal-cluster model. This is, in contrast to
traditional notions (minimum expenses or maximum profit), the sustainable
development of the crisis-free economical system, which corresponds
to extreme static fractal-cluster criteria $\left(D,\:\eta^{\Sigma},\:H\right)$
and obtained solutions for sustainable transformation (the criteria
of dynamical stability).

{}This generalized criterion of control optimization is a combination
of static and dynamical fractal-cluster criteria $\left(D,\:d^{2}H/dt^{2},\:\chi,\:\delta H,\:P\left[\delta^{2}H\right]\right)$,
and decisions on sustainable transformation of the micro-, meso- and
macro-level economic systems (\ref{f_t}--\ref{beta_sys}, \ref{f_t_exp}),
i.e., represents the conditions for sustainable crisis-free operation
of ES. However, this criterion does not ignore the traditional criteria
(minimum expenses and maximum profit, but enables the synergetic solution
to the problem of ES management optimization.

{}The criterion estimation of the resource utilization effectiveness
in economical system lead to two possibilities of the economic resource
controlling: 1) the resource redistribution in clusters for reaching
its ideal values $\left\{ \overline{C}_{i}^{ideal}\right\} $; 2)
the resource decreasing in next time period and the resource redistribution
in clusters for reaching their ideal values. The resource transformation
in clusters is realized on stable trajectories which is determined
by obtained solutions (\ref{f_t_eps}).

\begin{figure}[H]
{}\centering{}\includegraphics[scale=0.45]{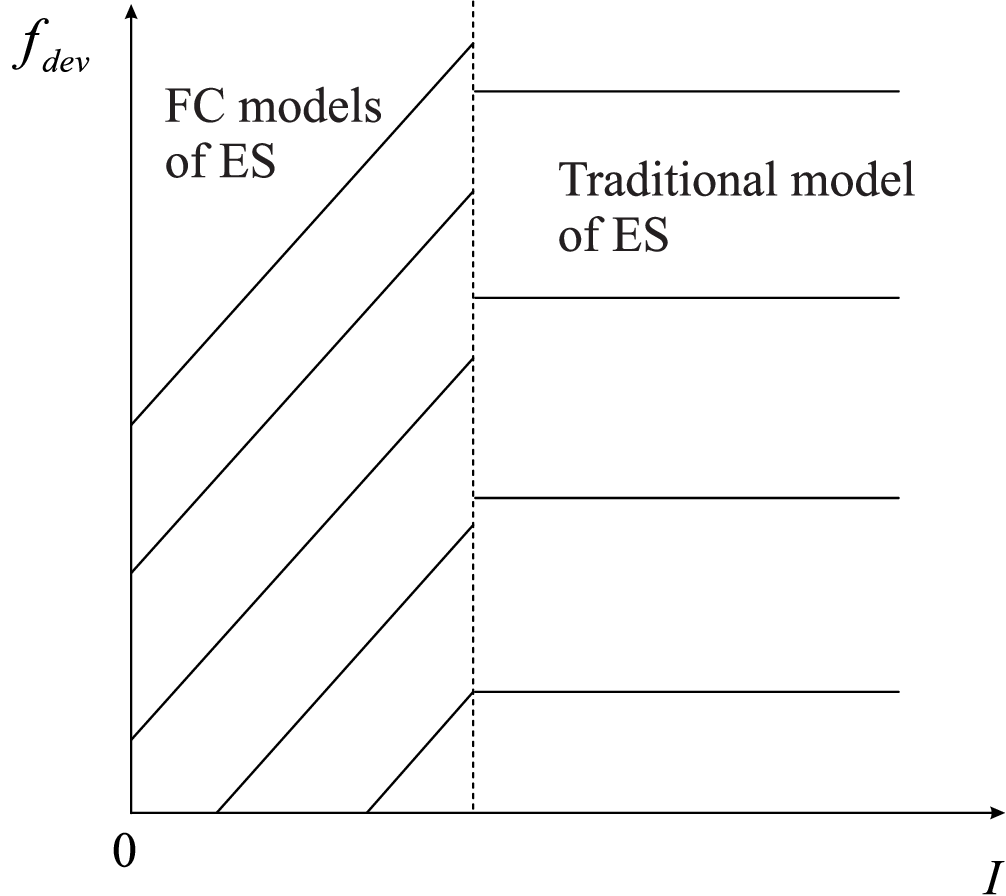}\caption{Possessing of ES economic \& mathematical models $I$ is an information
about output product's cost, $f_{dev}$ is a function of the economic
system's development. \label{figure14} }
\end{figure}

{}On \fig{figure14} we can see the comparative position of the
traditional and fractal-cluster methods initialization for the economic
systems. In the case of the output product (service) price minimum
of information of the one we have the fractal-cluster methods domination.
For example, a financial input in the social sphere, in education,
in fundamental science and in economic systems under crisis conditions,
suggests that fractal-cluster models will be more perspective. On
the contrary case, the traditional economic models (for example, the
Leontiev's models) will surpass the suggested models. In the latter
case, the fractal-cluster models will be an additional apparatus of
the economic analysis.

{}However, this class of models has its drawbacks: unconventional
approach -- in an explicit form, without additional empirical information,
it is impossible to determine the criteria of ES activities (revenue,
profit, profitability, etc.).

{}

{}A retrospective fractal-cluster analysis of municipal structures
management for the Moscow region and the Municipal Department of Nashua,
USA is presented in \tabl{tab4} as the first example. As\tabl{tab4}
shows, for the U.S. Municipal Department FCC is almost perfect, the
criterion of management effectiveness $D$ and the total system efficiency
is close to 100\%. For the municipal structures of the Moscow region
the most successful in terms of governance is the year of 1993.

{}The generalized criterion $\chi$ for Nashua city, USA is maximum,
which indicates optimal management.

\begin{table}[H]
{}\centering{}%
\begin{tabular}{|c|c|c|c|c|c|c|c|}
\hline
Structure name & Year & Entropy & Criterion & Efficiency & \multicolumn{3}{c|}{Relative deviation }\tabularnewline
 &  & $H$ & $D$ & $\eta^{^{\sum}}$ & \multicolumn{3}{c|}{from ideal}\tabularnewline
\cline{6-8} \cline{7-8} \cline{8-8}
 &  &  &  &  & $\varepsilon_{H}$ & $\varepsilon_{D}$ & $\varepsilon_{\eta}$\tabularnewline
\hline
Municipal structure & 1990 & 0.360 & 0.113 & 0.830 & 41.7\% & 88.6\% & 17\%\tabularnewline
\cline{2-8} \cline{3-8} \cline{4-8} \cline{5-8} \cline{6-8} \cline{7-8} \cline{8-8}
of the Moscow & 1993 & 0.564 & 0.876 & 0.969 & 8.7\% & 88.6\% & 3.1\%\tabularnewline
\cline{2-8} \cline{3-8} \cline{4-8} \cline{5-8} \cline{6-8} \cline{7-8} \cline{8-8}
region & 1996 & 0.407 & 0.723 & 0.926 & 3.4\% & 27.8\% & 7.5\%\tabularnewline
\hline
Municipal Department & 1993 & 0.603 & 0.970 & 0.990 & 2.42\% & 3\% & 1\%\tabularnewline
\cline{2-8} \cline{3-8} \cline{4-8} \cline{5-8} \cline{6-8} \cline{7-8} \cline{8-8}
of Nashua city, USA & 1994 & 0.616 & 0.988 & 0.996 & 0.4\% & 1.2\%  & 0.4\%\tabularnewline
\hline
\end{tabular}

$ $

\caption{Comparative analysis of municipal structures management.}
\label{tab4}
\end{table}

\begin{figure}[H]
{}\centering{}\includegraphics[scale=0.4]{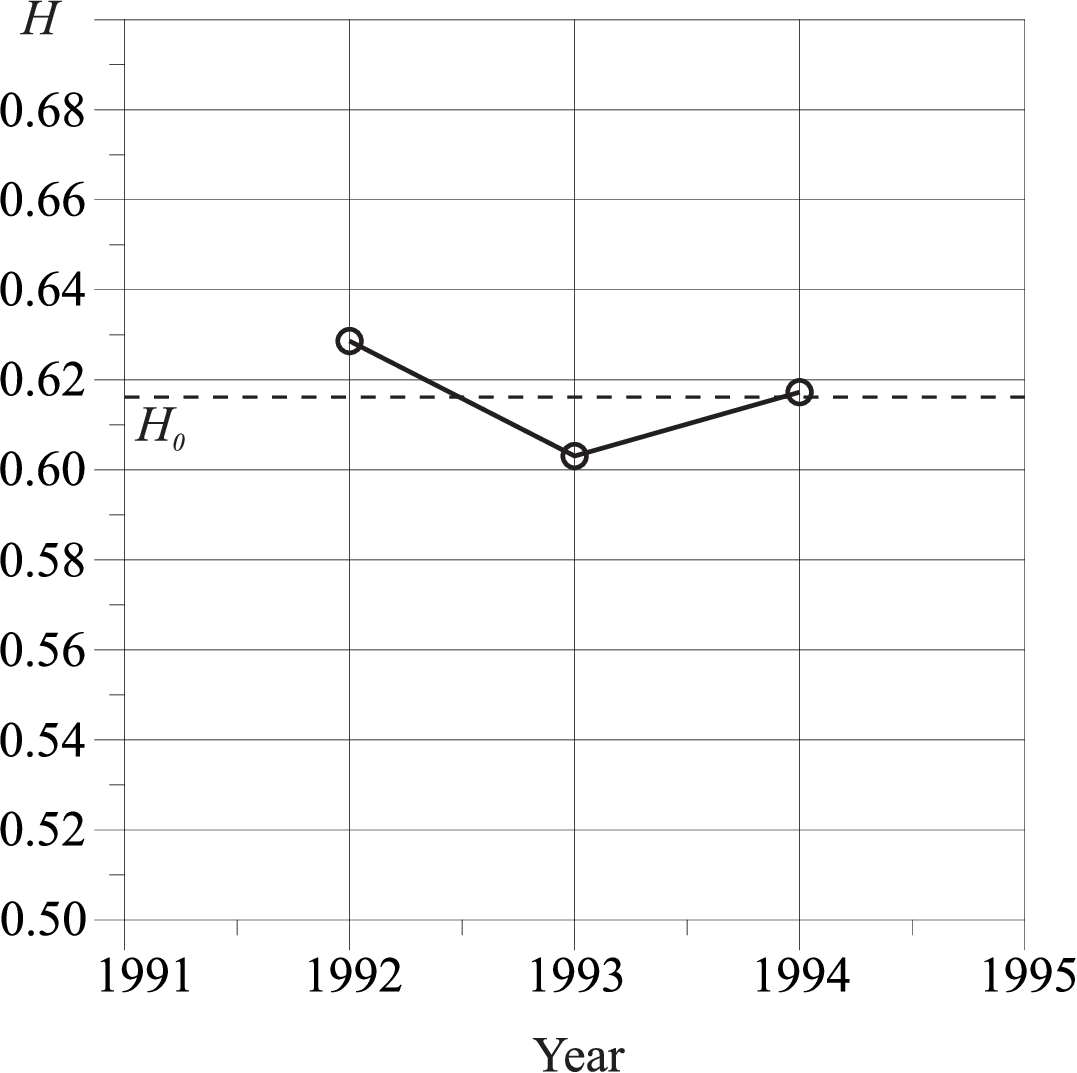}\caption{FC entropy oscilations near the \textquotedblleft Gold Ratio\textquotedblright .
\label{figure15} }
\end{figure}

{}In \fig{figure15} we can see the fractal-cluster entropy oscillations
near the ``Golden Ratio'' position. If the difference between the
oscillation of an entropy value and the ``Golden Ratio'' position
$\left(H_{0}\right)$will be essentially smaller than the ``Golden
Ratio'' value of the one, then we have the normal regime of the economical
system's functioning. In the countrary case we have the pathological
regime of the one.

{}

{}
\begin{figure}[H]
{}\centering{}\includegraphics[scale=0.8]{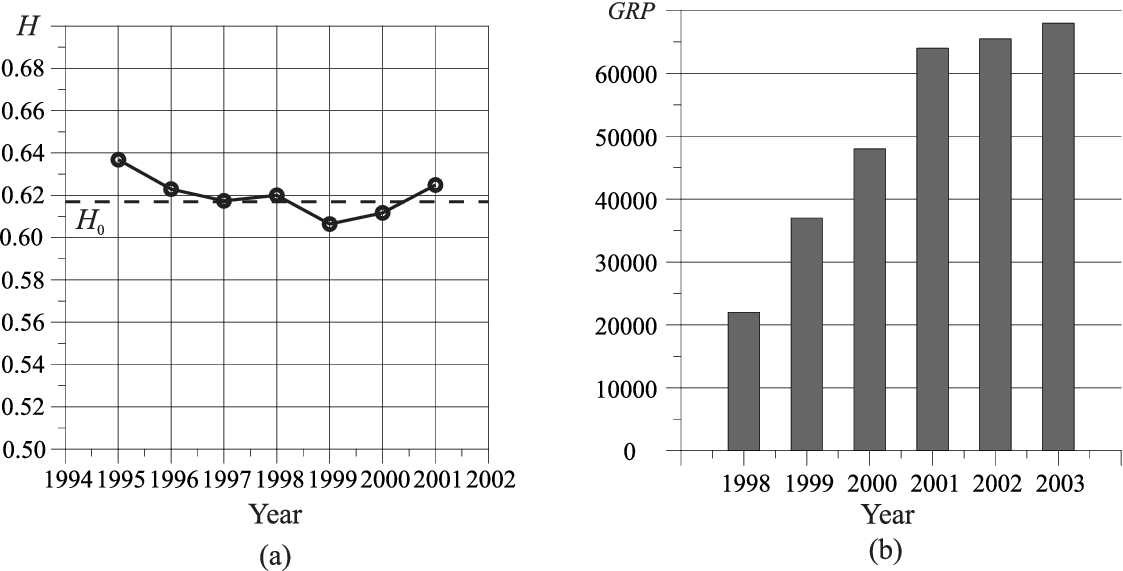}\caption{Changes in the FC entropy of the budget structure (a) and GRP (b)
for the period 1995--2001 in Samara region, Russia. \label{figure16} }
\end{figure}

{}The fractal-cluster analysis of the Samara region budget on the
data of 1995--2001 is presented in histograms (\fig{figure16}a).
From these illustrations it is seen that there is a weak oscillation
of the fractal cluster entropy around the value of $H_{0}\approx0.618$
is the ``Golden Ratio''$\left(\dfrac{\delta H}{H_{0}}\approx0.01\div0.03\right).$
This proves the efficient resource allocation of budget from 1995
to 2001. The results of fractal-cluster analysis are supported by
statistical data on the GRP, the pace of economical growth, investment,
rising of living standards in Samara region (\fig{figure16}b).

{}As can be seen on \fig{figure15} and \fig{figure16} there is
a confirmation of the hypothesis of low intensity waves $H$ (functional
instability) for the successful development of ES.

\begin{figure}[H]
{}\centering{}\includegraphics[scale=0.8]{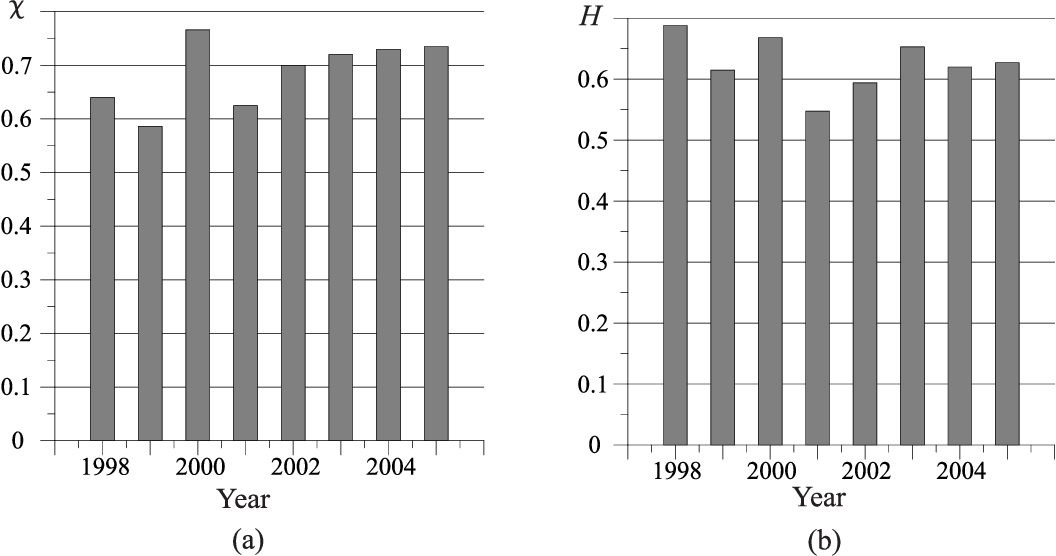}\caption{Histograms of the generalized criterion distribution (a), and the
fractal-cluster entropy distribution (b) in the financial structure
of Samara State University of Railway Engineering budget from 1998
to 2005. \label{figure17} }
\end{figure}

{}One more example illustrating the developed theory of resource
distribution in micro level ES is the result of analysis for Samara
State University of Railway Engineering. \fig{figure17} shows the
histograms of the generalized criterion $\chi\:$ (a) and fractal-cluster
entropy of the topological structure of the university budget $H\:$
(b) for Samara State University of Railway Engineering in the period
from 1998 to 2005. These histograms show clearly that the characteristics
of academy management $\left(H=0.688;\:D=0.593;\:\chi=0.64\right)$
were the worst in 1998, because of the general economical situation
of that time in the country; the consequences of default are noticeable
in the next 1999, respectively ($H=0.618$, $D=0.603$, and $\chi=0.5848$).

{}

{}It should be noted that despite the fact that the fractal-cluster
entropy distribution of the budget topological structure is close
to the ``Golden Ratio'' $\left(H_{0}=0.618\right)$, other criteria
$D$ and $\chi$are not equal to their maximum values $\left(D^{max}=\chi^{max}=1\right)$,
i.e. there is an imbalance in the structure of the academy budget.
This fact $\left(H=H_{0}\right)$ can be explained by the following:
with a scarce university budget and a high living wage (1250 rubles
per month in 1999), university authorities did not redistribute the
consolidated budget to increase wages, which could lead to an increase
entropy $H$.

{}In terms of structural budget management, the best in the academy
is the year of 2000 $\left(H=0.669;\:D=0.722;\:\chi=0.766\right)$.The
deterioration of consolidated budget management structure in 2001
is driven by objective reasons:

{}1) reduction of wage supplements from the Ministry of Transport,
which led to a decrease in the energy cluster, and thus to the entropy
$H$ decrease;

{}2) reconstruction of the university (construction of buildings,
dormitories), modernization and equipping of the laboratory-technical
base, financed by the Ministry of Transport and Kuibyshev Railway
led to a relative increase of technological cluster.

{}However, as the analysis shows, these endogenous components fluctuate
as parts of the university budget have a slight and transitory nature.

{}Important indicators of stability in the academy development are
a high integrated indicator of quality -- demand for graduates $~96$\%
(in Russia on average the figure is 50\%) and high rates of the university
employees' salary growth. Obviously, the fractal-cluster structuring
of the resources of the socio-economic system is an intra-system assessment
of the effectiveness of managing their resources.

{}To connect the characteristics of the fractal-cluster analysis
of the resources of the economic system with its valid integral characteristics
$v$ (GDP, labor productivity, capitalization, etc.), empirical dependencies
can be used in the following form: $v=F\left(D,\eta\right)$. It is
nessesary to emphasize that general process of the analysis for ES
has to enclude besides financial clusterization of the ES budget human's
clusterization and time budget clusterization.

\subsection{Biological organisms}

\begin{figure}[H]
{}\centering{}\centering{}\includegraphics[scale=0.9]{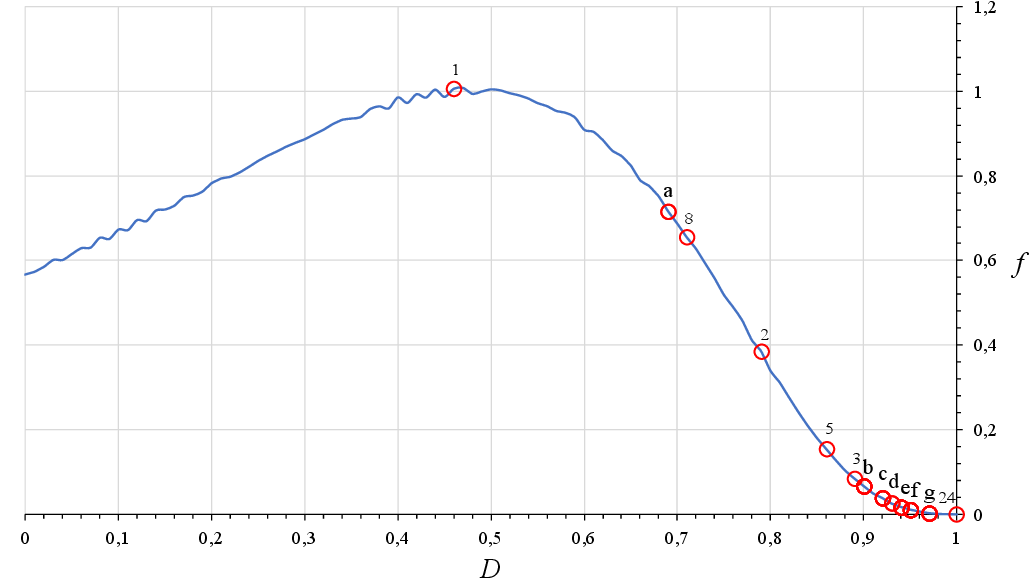}\caption{Stochastic FC law for the biological organisms. The circle numbers
in the figure correspond to the characteristics of the organisms in
Table \ref{tab6} (a -- 4, 16; b -- 7, 20, 31, 33, 38, 39; c --
26, 30, 32, 34, 35; d -- 11, 22; e -- 6, 9, 14, 15, 17, 19; f --
10, 12, 25, 27, 28, 36, 37; g -- 13, 18, 21, 23, 29). The vertical
axis represents the probability density of the organism's occurrence
and the horizontal axis shows values of the $D$-criterion. \label{figure18} }
\end{figure}

{}By using the mathematical tools of the fractal-cluster theory we
have determined the values of fractal-cluster entropy criteria $H$,
$F$-criterion and the effectiveness criterion $D$ for biological
organisms (\tabl{tab5} -- \tabl{tab6})). It has allowed us to
formulate three fractal-cluster laws: energy, evolutional and stochastic
for biological organisms:

{}1) the stochastic law, which determines the probability of appearance
of biological organisms (as shown in \fig{figure18}, there is a
bijection -- the most ancient biological organisms -- Hlamidomonas,
Hydra, had the highest probability of their occurrence);

\begin{figure}[H]
{}\centering{}\centering{}\includegraphics[scale=0.6]{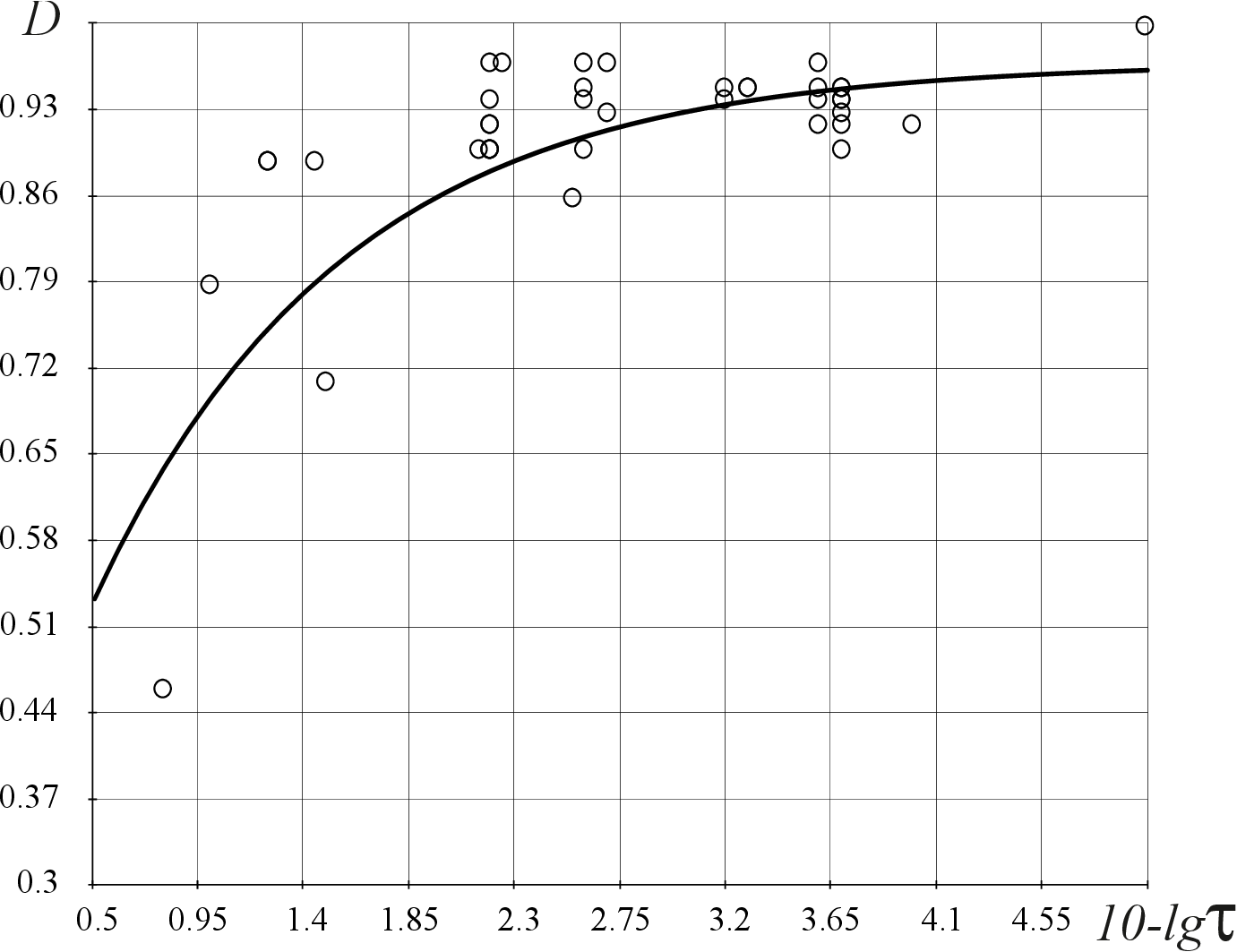}\caption{Evolutionary FC law for the biological organisms, $\tau$ -- is time
referes to the past, $D(\tau)\approx1-\dfrac{H_{0}}{2}t^{-\tfrac{1}{H_{0}}}$,
where $t=10-\lg\tau$, $H_{0}=0.618$, $\tau$ is the time of occurrence
organisms in the past (see, for example, \cite{Life_1,Life_2,Life_3,Life_4}
and references therein). The vertical axis represents values of the
$D$-criterion and the horizontal axis represents the function depending
on time $t=10-\lg\tau$. Circles in the figure correspond to different
organisms in the Table \ref{tab6}. \label{figure19} }
\end{figure}

{}2) the evolutional law (\fig{figure19}), illustrating the increasing
complexity and perfection of the emerging organisms;

\begin{figure}[H]
{}\centering{}\centering{}\includegraphics[scale=0.6]{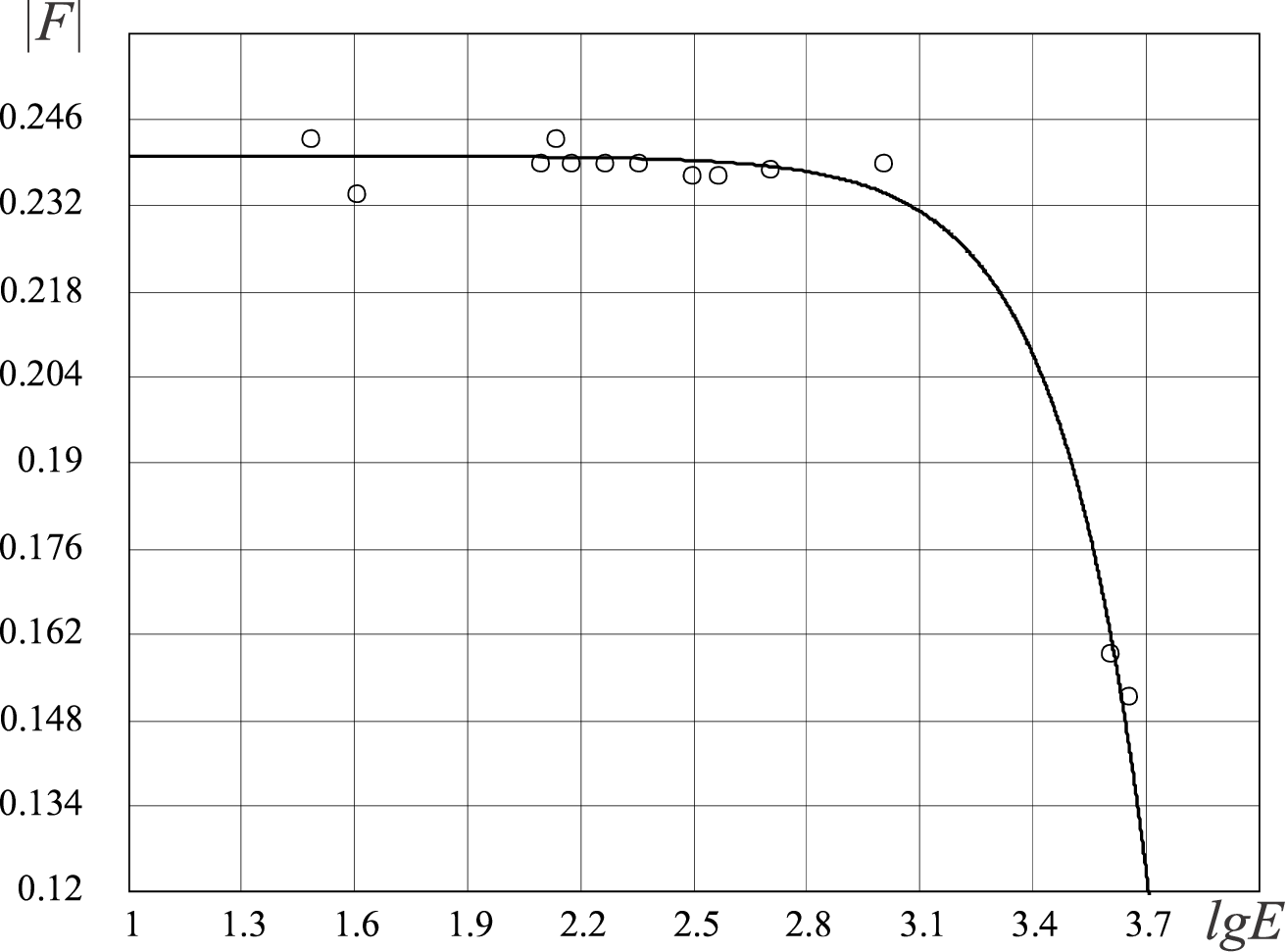}\caption{Energy FC law for the biological organisms, $F(E)\approx0.24-\dfrac{E^{3H_{0}}}{H_{0}\cdot10^{8}}$,
$H_{0}=0.618$. The vertical axis represents absolute values $\left|F\right|$
of the $F$-criterion and the horizontal axis represents values of
the log power consumption ($\lg E$). Circles in the figure correspond
to different organisms in the Table \ref{tab6}. \label{figure20}}

\end{figure}

{}3) the energy law (\fig{figure20}), which characterizes the energy
perfection of biological organisms (the dependence of the fractal-cluster
entropy $H$ or $F$-criterion on the energy consumption per $1$
$kg$ body weight per day, $E$).

{}The third law (the energy law) the is one that has a fundamental
correlation with the well-known problem -- of ``predator-prey'',
but the answer to the question formulated allows us to explain why
the predator cannot eat at all the victims. The answer to this question
is that the average value of the free fractal-cluster energy for a
predator and a victim will be equal.

{}The first fundamental fractal-cluster law has determined the probability
density of the biological organism origin.

{}From the trend first law, observed in we can see that the probability
density of the origin of ancient organisms is essentially bigger than
for the more modern organisms (\fig{figure18}). Evolution and energy
laws have the similar mathematical view:

{}
\begin{equation}
Z=A-\frac{Y^{a}}{\beta},\label{eq:42}
\end{equation}
where $Z=D,\:A=1,\:Y=\tau$ is for the evolutionary FC law, $Z=F,\:A=\ell,\:Y=E$
is for the energy FC law.

{}
\begin{table}[H]
{}\centering{}\caption{Cluster values for the biological organisms (males)}
\label{tab5} \vspace{1mm}

{}\centering{}%
\begin{tabular}{|p{0.1in}|p{2in}|>{\centering}p{0.5in}|>{\centering}p{0.35in}|>{\centering}p{0.35in}|>{\centering}p{0.35in}|>{\centering}p{0.35in}|>{\centering}p{0.35in}|}
\hline
 & {}Organism's names  & {}Mass, kg  & \multicolumn{5}{p{2.15in}|}{{}\centering FCR, \%}\tabularnewline
\hline
 &  &  & {}$\overline{C}_{1}$  & {}$\overline{C}_{2}$  & {}$\overline{C}_{3}$  & {}$\overline{C}_{4}$  & {}$\overline{C}_{5}$\tabularnewline
\hline
{}1  & {}Chlamidomonas  & {}$3\cdot10^{-11}$  & {}40$\pm$10  & {}10$\pm$8  & {}30$\pm$6  & {}10$\pm$5  & {}10$\pm$5\tabularnewline
\hline
{}2  & {}Hydra vulgaris  & {}$10^{-5}$  & {}40$\pm$10  & {}30$\pm$8  & {}10$\pm$6  & {}10$\pm$5  & {}10$\pm$5\tabularnewline
\hline
{}3  & {}Scorpiones mingrelicus  & {}$5\cdot10^{-4}$  & {}33$\pm$6  & {}33$\pm$5  & {}17$\pm$4  & {}9$\pm$3  & {}6$\pm$2\tabularnewline
\hline
{}4  & {}Oligochaeta  & {}$10^{-3}$  & {}19$\pm$8  & {}50$\pm$10  & {}16$\pm$4  & {}10$\pm$3  & {}5$\pm$2\tabularnewline
\hline
{}5  & {}Anisoptera libellula depressa  & {}$10^{-3}$  & {}40$\pm$8  & {}23$\pm$5  & {}17$\pm$4  & {}10$\pm$3  & {}10$\pm$2\tabularnewline
\hline
{}6  & {}Micromys minitus  & {}$5\cdot10^{-3}$  & {}40$\pm$6  & {}27$\pm$5  & {}16$\pm$4  & {}10$\pm$3  & {}7$\pm$2\tabularnewline
\hline
{}7  & {}Rona ridibunda  & {}0.05  & {}40$\pm$5  & {}30$\pm$4  & {}16$\pm$3  & {}8$\pm$2  & {}6$\pm$1\tabularnewline
\hline
{}8  & {}Testudo horsefieldi  & {}0.1  & {}38$\pm$6  & {}20$\pm$5  & {}30$\pm$4  & {}7$\pm$3  & {}5$\pm$1\tabularnewline
\hline
{}9  & {}Cucules canorus  & {}0.1  & {}40$\pm$6  & {}27$\pm$5  & {}16$\pm$4  & {}10$\pm$3  & {}7$\pm$2\tabularnewline
\hline
{}10  & {}Procellariida  & {}0.8  & {}40$\pm$6  & {}28$\pm$5  & {}16$\pm$4  & {}10$\pm$3  & {}6$\pm$2\tabularnewline
\hline
{}11  & {}Larus argentatus  & {}1.0  & {}39$\pm$6  & {}28$\pm$5  & {}16$\pm$4  & {}10$\pm$3  & {}7$\pm$2\tabularnewline
\hline
{}12  & {}Heroestes edwardsi  & {}3.0  & {}40$\pm$6  & {}27$\pm$5  & {}17$\pm$4  & {}10$\pm$3  & {}6$\pm$2\tabularnewline
\hline
{}13  & {}Ciconia ciconia  & {}4.0  & {}40$\pm$6  & {}27$\pm$5  & {}16$\pm$4  & {}11$\pm$3  & {}6$\pm$2\tabularnewline
\hline
{}14  & {}Lepus timidus  & {}5.0  & {}40$\pm$6  & {}28$\pm$5  & {}16$\pm$4  & {}11$\pm$3  & {}5$\pm$2\tabularnewline
\hline
{}15  & {}Grus grus  & {}6.0  & {}40$\pm$6  & {}27$\pm$5  & {}16$\pm$4  & {}10$\pm$3  & {}7$\pm$2\tabularnewline
\hline
{}16  & {}Paralithodes camtchatica  & {}7.0  & {}20$\pm$6  & {}50$\pm$5  & {}16$\pm$4  & {}8$\pm$3  & {}6$\pm$2\tabularnewline
\hline
{}17  & {}Pelecanida onocrotalus  & {}10  & {}40$\pm$6  & {}28$\pm$5  & {}16$\pm$4  & {}11$\pm$3  & {}5$\pm$2\tabularnewline
\hline
{}18  & {}Vulpes  & {}10  & {}40$\pm$6  & {}27$\pm$5  & {}16$\pm$4  & {}11$\pm$3  & {}6$\pm$2\tabularnewline
\hline
{}19  & {}Castor fiber  & {}30  & {}40$\pm$6  & {}26$\pm$5  & {}17$\pm$4  & {}12$\pm$3  & {}5$\pm$2\tabularnewline
\hline
{}20  & {}Acinonyx jubatus  & {}50  & {}40$\pm$6  & {}30$\pm$5  & {}16$\pm$4  & {}8$\pm$3  & {}6$\pm$2\tabularnewline
\hline
{}21  & {}Canis lipus  & {}50  & {}40$\pm$6  & {}27$\pm$5  & {}16$\pm$3  & {}11$\pm$3  & {}6$\pm$2\tabularnewline
\hline
{}22  & {}Pan troglodytes  & {}60  & {}39$\pm$6  & {}28$\pm$5  & {}16$\pm$3  & {}11$\pm$2  & {}6$\pm$2\tabularnewline
\hline
{}23  & {}Orycturopus afer  & {}70  & {}40$\pm$6  & {}28$\pm$5  & {}17$\pm$3  & {}11$\pm$2  & {}5$\pm$2\tabularnewline
\hline
{}24  & {}Homo sapiens  & {}75  & {}38$\pm$6  & {}27$\pm$5  & {}16$\pm$4  & {}13$\pm$2  & {}6$\pm$1\tabularnewline
\hline
{}25  & {}Ursus arctos  & {}100  & {}40$\pm$6  & {}27$\pm$5  & {}17$\pm$4  & {}10$\pm$3  & {}6$\pm$2\tabularnewline
\hline
{}26  & {}Cervina nippon  & {}120  & {}40$\pm$6  & {}28$\pm$5  & {}17$\pm$4  & {}9$\pm$3  & {}6$\pm$2\tabularnewline
\hline
{}27  & {}Sus scrofa  & {}150  & {}40$\pm$6  & {}28$\pm$5  & {}16$\pm$4  & {}10$\pm$3  & {}6$\pm$2\tabularnewline
\hline
{}28  & {}Pongo pygmaeus  & {}200  & {}39$\pm$6  & {}29$\pm$5  & {}16$\pm$3  & {}10$\pm$3  & {}6$\pm$2\tabularnewline
\hline
{}29  & {}Gorilla gorilla  & {}250  & {}39$\pm$6  & {}28$\pm$5  & {}16$\pm$3  & {}11$\pm$2  & {}6$\pm$1\tabularnewline
\hline
{}30  & {}Equida burchelli  & {}300  & {}40$\pm$7  & {}28$\pm$6  & {}17$\pm$4  & {}9$\pm$3  & {}6$\pm$2\tabularnewline
\hline
{}31  & {}Tursiops  & {}400  & {}42$\pm$7  & {}28$\pm$6  & {}16$\pm$4  & {}8$\pm$3  & {}6$\pm$2\tabularnewline
\hline
{}32  & {}Equus caballus  & {}400  & {}40$\pm$7  & {}28$\pm$6  & {}17$\pm$4  & {}9$\pm$3  & {}6$\pm$2\tabularnewline
\hline
{}33  & {}Galeocerdo cuvieri  & {}500  & {}40$\pm$6  & {}30$\pm$5  & {}16$\pm$4  & {}8$\pm$3  & {}6$\pm$2\tabularnewline
\hline
{}34  & {}Camelus bactrianus  & {}600  & {}40$\pm$6  & {}27$\pm$5  & {}18$\pm$4  & {}9$\pm$3  & {}6$\pm$2\tabularnewline
\hline
{}35  & {}Giraffa cameleopardalis  & {}750  & {}40$\pm$6  & {}29$\pm$5  & {}16$\pm$4  & {}10$\pm$3  & {}5$\pm$2\tabularnewline
\hline
{}36  & {}Hippopotamus amphibius  & {}3000  & {}40$\pm$6  & {}28$\pm$5  & {}16$\pm$4  & {}10$\pm$3  & {}6$\pm$2\tabularnewline
\hline
{}37  & {}Loxodonta africana  & {}5000  & {}40$\pm$6  & {}28$\pm$5  & {}16$\pm$4  & {}10$\pm$3  & {}6$\pm$2\tabularnewline
\hline
{}38  & {}Balaena mysticetus  & {}150000  & {}42$\pm$5  & {}28$\pm$4  & {}16$\pm$3  & {}8$\pm$2  & {}6$\pm$1\tabularnewline
\hline
{}39  & {}Balaenoptera musculus  & {}200000  & {}42$\pm$5  & {}28$\pm$4  & {}16$\pm$3  & {}8$\pm$2  & {}6$\pm$1\tabularnewline
\hline
\end{tabular}
\end{table}

{}
\begin{table}[H]
{}\centering{}\centering{}\caption{FC criterion estimation for the biological organisms}
\label{tab6} \vspace{1mm}
\begin{tabular}{|c|c|c|c|c|c|}
\hline
 & {}Organism's names  & {}$D$  & {}$H$  & {}$\overline{C}_{1}$  & {}$F$\tabularnewline
\hline
{}1  & {}Chlamidomonas  & {}0.46  & {}0.622  & {}0.387  & {}0.235\tabularnewline
\hline
{}2  & {}Hydra vulgaris  & {}0.79  & {}0.637  & {}0.4  & {}0.237\tabularnewline
\hline
{}3  & {}Scorpiones mingrelicus  & {}0.89  & {}0.561  & {}0.338  & {}0.223\tabularnewline
\hline
{}4  & {}Oligochaeta  & {}0.69  & {}0.342  & {}0.19  & {}0.152\tabularnewline
\hline
{}5  & {}Anisoptera libellula depressa  & {}0.86  & {}0.832  & {}0.4  & {}0.238\tabularnewline
\hline
{}6  & {}Micromys minitus  & {}0.94  & {}0.639  & {}0.4  & {}0.239\tabularnewline
\hline
{}7  & {}Rona ridibunda  & {}0.90  & {}0.639  & {}0.4  & {}0.239\tabularnewline
\hline
{}8  & {}Testudo horsefieldi  & {}0.71  & {}0.614  & {}0.38  & {}0.234\tabularnewline
\hline
{}9  & {}Cucules canorus  & {}0.94  & {}0.639  & {}0.4  & {}0.239\tabularnewline
\hline
{}10  & {}Procellariida  & {}0.95  & {}0.662  & {}0.42  & {}0.242\tabularnewline
\hline
{}11  & {}Larus argentatus  & {}0.93  & {}0.627  & {}0.39  & {}0.237\tabularnewline
\hline
{}12  & {}Heroestes edwardsi  & {}0.95  & {}0.639  & {}0.4  & {}0.239\tabularnewline
\hline
{}13  & {}Ciconia ciconia  & {}0.97  & {}0.664  & {}0.421  & {}0.243\tabularnewline
\hline
{}14  & {}Lepus timidus  & {}0.94  & {}0.639  & {}0.4  & {}0.239\tabularnewline
\hline
{}15  & {}Grus grus  & {}0.94  & {}0.664  & {}0.421  & {}0.243\tabularnewline
\hline
{}16  & {}Paralithodes camtchatica  & {}0.69  & {}0.359  & {}0.2  & {}0.159\tabularnewline
\hline
{}17  & {}Pelecanida onocrotalus  & {}0.94  & {}0.639  & {}0.4  & {}0.239\tabularnewline
\hline
{}18  & {}Vulpes  & {}0.97  & {}0.639  & {}0.4  & {}0.239\tabularnewline
\hline
{}19  & {}Castor fiber  & {}0.94  & {}0.664  & {}0.421  & {}0.243\tabularnewline
\hline
{}20  & {}Acinonyx jubatus  & {}0.90  & {}0.639  & {}0.4  & {}0.239\tabularnewline
\hline
{}21  & {}Canis lipus  & {}0.97  & {}0.639  & {}0.4  & {}0.239\tabularnewline
\hline
{}22  & {}Pan troglodytes  & {}0.93  & {}0.627  & {}0.39  & {}0.237\tabularnewline
\hline
{}23  & {}Orycturopus afer  & {}0.97  & {}0.633  & {}0.395  & {}0.238\tabularnewline
\hline
{}24  & {}Homo sapiens  & {}1.0  & {}0.614  & {}0.38  & {}0.234\tabularnewline
\hline
{}25  & {}Ursus arctos  & {}0.95  & {}0.639  & {}0.4  & {}0.239\tabularnewline
\hline
{}26  & {}Cervina nippon  & {}0.92  & {}0.639  & {}0.4  & {}0.239\tabularnewline
\hline
{}27  & {}Sus scrofa  & {}0.95  & {}0.639  & {}0.4  & {}0.239\tabularnewline
\hline
{}28  & {}Pongo pygmaeus  & {}0.95  & {}0.627  & {}0.39  & {}0.237\tabularnewline
\hline
{}29  & {}Gorilla gorilla  & {}0.97  & {}0.627  & {}0.39  & {}0.237\tabularnewline
\hline
{}30  & {}Equida burchelli  & {}0.92  & {}0.636  & {}0.4  & {}0.239\tabularnewline
\hline
{}31  & {}Tursiops  & {}0.90  & {}0.659  & {}0.42  & {}0.242\tabularnewline
\hline
{}32  & {}Equus caballus  & {}0.92  & {}0.636  & {}0.4  & {}0.239\tabularnewline
\hline
{}33  & {}Galeocerdo cuvieri  & {}0.90  & {}0.639  & {}0.4  & {}0.239\tabularnewline
\hline
{}34  & {}Camelus bactrianus  & {}0.92  & {}0.639  & {}0.4  & {}0.239\tabularnewline
\hline
{}35  & {}Giraffa cameleopardalis  & {}0.92  & {}0.639  & {}0.4  & {}0.239\tabularnewline
\hline
{}36  & {}Hippopotamus amphibius  & {}0.95  & {}0.639  & {}0.4  & {}0.239\tabularnewline
\hline
{}37  & {}Loxodonta africana  & {}0.95  & {}0.639  & {}0.4  & {}0.239\tabularnewline
\hline
{}38  & {}Balaena mysticetus  & {}0.90  & {}0.663  & {}0.42  & {}0.243\tabularnewline
\hline
{}39  & {}Balaenoptera musculus  & {}0.90  & {}0.663  & {}0.42  & {}0.243\tabularnewline
\hline
\end{tabular}
\end{table}

\newpage{}

{}

{}

{}

\section{Conclusions}

{}\label{Concl}

{}The general basis of the fractal-cluster theory for the analysis
and control of self-organizing systems has been presented. The first
part of the determined theory includes the criterion's apparatus of
the resource distribution in self-organizing systems of different
nature. The second part of the theory is based on the probable apparatus
of the fractal-cluster system's researching, which includes Kolmogorov--Feller
equation for the configuration probability and stationary solution
of the one on a numerical research basis. The calculated values of
the mathematical expectation of combinations of clusters $\overline{C}_{i}$,
as well as their most probable values in the fractal-cluster space,
represent a probabilistic confirmation of the values of the reference
(ideal) values of clusters $\overline{C}_{i}^{ideal}$ obtained experimentally
by V.P. Burdakov \cite{Burd}.

{}Testing of the FC theory for the biological organisms' research
was realized on the basis of study \cite{Burd,Burd_Th,Volov_Entropy}
(\tabl{tab5}). It should be noted that the probabilistic fractal-cluster
law may have a different thermodynamic interpretation. For example,
the vector $\overrightarrow{W}$ representing the conditional function
of the development of organisms can be described by an analogue of
Fourier's law:
\[
\overrightarrow{W}=-a\mathrm{\cdot\overrightarrow{grad}}f,
\]
where $f$ is the probability density of cluster combinations $\overline{C}_{i}$.
Nevertheless, this hypothesis requires a separate study. In the result
of the analysis based on the FC approach three fundamental FC laws
for biological organisms were identified (\cite{Volov5,Volov6}, \tabl{tab6}):

{}1) the FC stochastic law defines the density of probability of
occurrence of biological organisms depending on the perfection of
the resource allocation in the organism -- the $D$-criterion (\fig{figure18})
shows that there is a bijection -- the simpler biological organisms
(the $D$-criterion less than perfect) -- Chlamidomonada, Hydra,
had the highest probability density of its appearance);

{}2) the FC evolutionary law (\fig{figure19} ) illustrates the
increasing complexity and perfection of the emerging organisms through
the time, from the simpler organisms and to humans;

{}3) the FC power law (\fig{figure20}) characterizes the energy
perfection of biological organisms (the dependence of the FC entropy
or the $F$-criterion of the energy consumption per $1$ kg of organism's
weight per day (\fig{figure19} ).

{}The stochastic law of development of biological organisms is obtained
due to the distribution of possible states based on the $D$-criterion
and correlates this distribution with the corresponding values of
the $D$-criterion for biological organisms (\tabl{tab5}). The second
FC law shows the growth of perfection of the resource distribution
in organisms (the $D$-criterion) over the time (\fig{figure19}).
\fig{figure20} shows the third revealed FC law that defines a link
between the FC free energy of biological organisms and the level of
energy consumption per 1 kg of weight. This law represents, in a certain
sense, the thermodynamic response to the output of Lotka--Volterra's
problem (``predator-prey'', \fig{figure1}) judging by the level
of the FC free energy ($F$) of biological organisms (including mammals,
fish, insects, etc., with the exception of worms -- the last right
point in \fig{figure20} -- it has on average the same value, i.e
it has an average energy balance between the ``predators'' and their
victims - the ``predator'' cannot catch the ``victim''. The energy
law for biological organisms (\fig{figure20}) is a thermodynamic
confirmation of decision of the problem -- ``predator-prey'', ``predators''
cannot kill all the potential ``victims'' (\fig{figure20}), i.e.
only the ``victim'' having a level of free energy is lower than
its average in the population will die. It can be summarized that
the decision of the ``pendulum'' of Lotka--Volterra is a dynamic
energy condition of impossibility of the destruction the ``victims''
by the ``predactors'' and the FC energy law is the static energy
condition of the one. It is important to note that the mathematical
expression for the second and third laws has the same view (\ref{eq:42}).

{}The review of biophysics studies (see Section \ref{Introduction})
shows that the current stage of convergence of biology and physics
reflects an increase in the number of publications which use the apparatus
of nonequilibrium thermodynamics of Prigogine. As shown, there are
researches which go beyond the traditional approach to this issue,
for example, Burdakov's study may be related to these researches.
Based on Burdakov's research \cite{Burd} for organisms and the analytical
apparatus of Prigogine's theory \cite{PN} the foundations of the
fractal-cluster theory for the analysis of organisms have been presented.
As known, Haken's approach allows us to describe the organism's trajectory
in phase space at a known value of the ``order parameter'' which
is determined from the experiment. Prigogine's approach does not need
experimental data for the description of organism's functioning as
the one has the universal thermodynamic tools at its basis. But Prigogine's
approach cannot describe the trajectory of an organism. The article
suggests the new alternative approach to systematization and the study
of the evolution of biological organisms on the basis of the fractal-cluster
theory. This theory in the certain sense presents the compromise between
Prigogine's \cite{PN,GP}and Haken's approaches \cite{Haken} in synergetics
and allows us to describe the organism phase trajectory in the $5^{n}$-dimensional
fractal-cluster space without additional experimental information.
It presents the criterion's apparatus of the resource distribution
in organisms and the stochastic distributions for the organism states
in the $5^{n}$-dimensional FC space.

{}The article gives a definition of the organism in the fractal-cluster
approach which is qualitatively different from the generally accepted
in biology and biophysics. But, at the same time, it shows the relationship
of seven classic factors of the organism's definition \cite{Campbell}
and the definition of the organism with the FC approach. The article
also proves that von Stockar's findings \cite{VonStockar2,VonStockar3,Campbell}
of the optimal growth efficiency for organisms have connection with
the fractal-cluster theory. In agreement with the FC theory (see Section
\ref{Criteria}, eq. (\ref{F})) the free energy is determined as
following:

{}
\[
F=\left|\overline{C}_{1}-H\right|,
\]
where $\overline{C}_{1}$ is the energy cluster value, $H$ is the
fractal-cluster entropy. For the two-dimensional symmetric FC matrix
the free energy $F$ is equal $F=\left|-\overline{C}_{1}+\overline{C}_{1}^{2}\right|$.
The maximal absolute value of the free energy is defined as the following:

{}
\[
\dfrac{dF}{d\overline{C}_{1}}=0,\;\overline{C}_{1}=0.5,\;F_{\max}=0.25.
\]
The optimal value of the free energy corresponds to ideal (standard)
values of clusters and is defined as the following:

{}
\[
\overline{C}_{1}^{(\mathrm{ideal})}=0.38,\;F_{\mathrm{opt}}=0.235.
\]
Thus, the optimal value of the free energy is 94\% from the maximal
value of the one:

{}
\[
F_{\mathrm{opt}}=0.94F_{\max}.
\]

{}Ergo, the share of free energy going to the cell growth will be
smaller than $F_{\mathrm{opt}}$: $F_{\mathrm{grows}}<F_{\mathrm{opt}}$
i.e smaller 100 \%. This fact qualitatively and quantitatively confirms
von Stockar's result by the FC theory. As a sample of the FC tool
using for the organism's analysis the FC criterion estimation for
the systematization of biological organisms has been presented. The
conducted criterion analysis of 39 species of biological organisms
allowed us to identify three fundamental FC laws: 1) the FC stochastic
law; 2) the FC evolutionary law; 3) the FC energy law. The probabilistic
law connects the thermodynamic perfection of the organism (the $D$-criterion)
with the fractal-cluster density of probability and the time of the
appearance of organism species: the simpler FC organisms (the $D$-criterion
is less) have a greater likelihood of their occurrence. The evolutionary
law connects the thermodynamic perfection of the organism with the
time of its occurrence.

{}The fractal-cluster theory is an additional toolkit to the generally
accepted theory of the evolution of organisms, which makes it possible
to give an exclusively thermodynamic assessment of their development
and functioning. In particular, an alternative solution to the famous
Volterra-Lotka's problem is presented for the first time on the basis
of fractal-cluster criteria: a static energy assessment was obtained
about the impossibility of complete destruction of the ``prey''
by the ``predator'', in contrast to the known dynamic solution of
this problem. There is doubtless that the new analytical apparatus
allows us to obtain a new knowledge about the object under study.

{}As for the prospects of applications of the fractal-cluster theory
in biology, they probably can be implemented in cell technologies,
in virology, as well as in medico-biological applications. The impact
of the external environment on the organism leads to a redistribution
of the organism's resources, which, in turn, makes it possible to
use the fractal-cluster criteria and, as a result, to obtain characteristics
of the stability of the organism's functioning, an opportunity to
obtain information in advance about the pathological trends of its
development.

{}For the economical systems the developed fractal-cluster theory
allows us to analyze and optimize SOS in the aspect of resource distribution.
A new generalized criterion of ES management optimization is formulated
on the basis of the fractal-cluster criteria and solutions for sustainable
transformation that have been developed. Testing of the developed
theory of analyzing resource distribution in the micro-, meso- and
macro-level ES affirmed its foundations and recommendations. The fractal-cluster
theory and models worked out on its basis will be dominant in the
prognosis of ES development, where it is impossible or difficult to
evaluate the system product in terms of value (educational institutions,
fundamental science, etc.). For ES ``inputs -- output'', with a
high degree of probability while predicting the products and assessing
the ES effectiveness (profitability, sales, GDP, etc.), traditional
economic and mathematical models will dominate. For this class of
ES the proposed fractal-cluster theory and models based on it, can
be used as auxiliary tools of ES management analysis

{}The developed fractal-cluster theory allows us to analyze and optimize
SOS in the aspect of resource distribution. A new generalized criterion
of ES management optimization is formulated on the basis of the fractal-cluster
criteria and solutions for sustainable transformation that have been
developed. Testing of the developed theory of analyzing resource distribution
in the micro-, meso- and macro-level ES affirmed its foundations and
recommendations.

{}The fractal-cluster theory and models worked out on its basis will
be dominant in the prognosis of ES development, where it is impossible
or difficult to evaluate the system product in terms of value (educational
institutions, fundamental science, etc.). For ES ``inputs - output'',
with a high degree of probability while predicting the products and
assessing the ES effectiveness (profitability, sales, GDP, etc.),
traditional economic and mathematical models will dominate. For this
class of ES the proposed fractal-cluster theory and models based on
it, can be used as auxiliary tools of ES management analysis (\fig{figure14}).

{}In the author's opinion, future researching and applications of
the fractal-cluster theory can be realized in the following ways:

{}1) obviously, the fractal-cluster fundamental research will be
connected with stochastic dynamics of the fractal-cluster systems;

{}2) the fractal-cluster theory can find a wide spectrum of use in
economic applications (a risk estimation in bank business, industry,
financial business etc.);

{}3) social-economic researches in science, the social sphere, and
education;

{}4) fundamental researches in biology.

{}\medskip{}

{}Acknowledgments

{}This work was partially supported by Russian Foundation for Basic
Research grant (project 13-01-00790-a).

{}I would like to thank prof. V. Burdakov and prof. V. Makarov for
usefull advices and important comments concerning the present article.
My especial thank to prof. A. Zubarev and A. Prisker for usefull discussions
and preparing this article.

\end{document}